\documentclass[preprint]{aastex}

\def \degpoint {${.}\!\!^{\circ}\!$}
\def \secpoint {${.}\!\!^{\prime \prime}\!$}
\def \minpoint {${.}\!\!^{\prime}$}

\shorttitle{Outer Halo of M31}
\shortauthors{Durrell et al.}

\begin{document}

\title{Photometry and the Metallicity Distribution of the Outer Halo of M31}

\author{Patrick R. Durrell\altaffilmark{1}}
\affil{Department of Physics \& Astronomy, 2219 Main Mall, University of British Columbia,
    Vancouver, BC   V6T 1Z4 CANADA }

\and

\affil{Department of Astronomy \& Astrophysics, The Pennsylvania State University, 525 Davey Lab, 
University Park, PA  16802  USA\altaffilmark{2}}
\email{pdurrell@astro.psu.ca}

\author{William E. Harris\altaffilmark{1}}
\affil{Department of Physics \& Astronomy, McMaster University, Hamilton, ON  L8S 4M1  CANADA}
\email{harris@physics.mcmaster.ca}

\and

\author{Christopher J. Pritchet}
\affil{Department of Physics \& Astronomy, University of Victoria, Victoria, BC  V8W 3P6  CANADA}
\email{pritchet@clam.phys.uvic.ca}

\altaffiltext{1}{Visiting Astronomer, Canada-France-Hawaii Telescope, operated by the National Research Council of Canada, le Centre National de la Recherche Scientifique de France, and the University of Hawaii}

\altaffiltext{2}{current address}

\begin{abstract}

We have conducted a wide-field CCD-mosaic study of the resolved red-giant branch stars 
of M31, in a field located 20 kpc from the nucleus along the SE minor axis.  In our 
($I, V-I$) color-magnitude diagram, red-giant branch (RGB) stars in the top three magnitudes
of the M31 halo are strongly present.  We use photometry of a more distant control field to
subtract field contamination and then to derive the ``cleaned'' luminosity function and metallicity
distribution for this outer-halo region of M31.  From the color 
distribution of the foreground Milky Way 
halo stars, we find a reddening $E(V-I)= 0.10 \pm 0.02$ for this field, and
from the luminosity of the RGB tip, we determine  a distance 
modulus $(m-M)_o = 24.47 \pm 0.12$ (= $783 \pm 43$ kpc).
The metallicity distribution function (MDF) is derived from interpolation within
an extensive new grid of RGB models (Vandenberg et al.~2000).
We find that the MDF is dominated by a moderately high-metallicity population 
([m/H]$\sim -0.5$) that has previously been found 
in more interior M31 halo/bulge fields, and is very much more metal-rich than the [m/H] $\sim -1.5$
level which characterizes the Milky Way halo.  In addition, a significant 
($\sim$30\% $-$ 40\%, depending on AGB star contribution) 
metal-poor population is also present.  
To first order, the total shape of the MDF resembles that 
predicted by a simple, single-component model of chemical evolution starting from primordial gas
with an effective  yield $y=0.0055$.  It strongly resembles the MDF recently found for the outer
halo of the giant elliptical NGC 5128 (Harris et al.~2000), though NGC 5128 has an even lower 
fraction of low-metallicity stars.  Intriguingly, in both NGC 5128 and M31, the metallicity distribution
of the {\it globular clusters} in M31 does not match the halo {\it stars}, in the sense that 
the clusters are far more heavily weighted to metal-poor objects.
We suggest similarities in the formation and early evolution of massive, spheroidal stellar systems.

\end{abstract}



\section{Introduction}
 
While the stellar halo may be but a minor constituent 
in the total mass and luminosity of a galaxy, it is an important tracer of the 
conditions of the formation and earliest evolution of 
galaxies.  Studies of the halo stars and globular clusters in our own Milky Way have 
provided a wealth of insight into the early chemical evolution of the Galaxy, 
including its use as a chronometer 
to probe formation timescales and models.    
Our location within the MW, however, makes the study of more global 
properties of our halo (such as its radial extent, age distribution, 
chemical composition -- all of which require large, complete datasets) very difficult,
since in a given star field there are far more local disk stars than halo stars, 
usually by over 3 orders of magnitude.   Monumental 
`needle in a haystack' efforts to both find and study 
MW halo stars have been made \citep[eg.][and references within]{beer96, car96, som97}; 
and further ambitious studies are currently underway \citep{tot98, maj00, mor00, iv00}.    

A different, yet equally interesting, approach is to study 
the halo stars in nearby galaxies where the stars are at a 
common distance, relieving one of the primary difficulties 
with studying most MW halo populations.  The halo of the spiral galaxy 
M31 provides us with a nearby and readily accessible population of
large numbers of halo stars; photometry of its 
red-giant branch (RGB) stars down to the level of the horizontal 
branch is easily carried out with 4-meter-class 
telescopes.   M31 is also of earlier Hubble type (Sb vs. SBbc) 
and more massive than the Milky Way 
\citep[with a higher proportion in the bulge/spheroid component, eg.][and 
references within; but see Evans \& Wilkinson 2000]{cvdb99, free99, cote00b}.
Unlike the MW bulge, the larger bulge of M31 follows 
an $r^{1/4}$ profile \citep[][hereafter PvdB94]{pv94}, and thus it 
is of interest to understand the bulge and halo in both galaxies 
to investigate any contrasts in formation mechanisms \citep[see also][]{wyse97, mor99}.
We will find that the M31 halo stars are indeed strikingly different on
average from those in the Milky Way.

While spectroscopy of stars in M31 is difficult and time-consuming 
\citep[see][for some first results]{reit00}, it
is not hard to obtain photometry of the RGB stars through 
metallicity-sensitive indices to study the metallicity
distribution function (MDF) in its broad terms.  \citet{mk86}
were the first to clearly resolve the RGB stars, in a field 7 kpc from the
galaxy center along the southern minor axis.  Their data indicated that the M31 halo 
there has [Fe/H]$\sim -0.6$, more metal-rich by a factor of a few
than the MW halo.    \citet{mou86} performed a similar study on fields located 
5, 12 and 20 kpc from M31 along the minor axis; analysis of the 12 kpc field 
showed a similarly high metallicity, but contamination made results from the other 2 
fields less certain.    Deeper ground-based studies 
at other locations in the inner M31 halo ($r_{M31} \sim 7 - 12$ kpc) 
confirmed the high metallicity for 
most of the M31 halo population \citep{pv88, ch91, dav93, dur94, cou95}. 
Subsequent HST-based CMDs yielded measurements reaching deep enough to show
the horizontal-branch stars in its halo clusters and field
\citep{hol96, rich96a, rich96b}, and these convincingly showed that 
a metal-poor component does indeed 
exist \citep[which had already been suggested by the presence 
of RR Lyrae stars;][]{pv87}.  Recent photometric and spectroscopic work 
by \citet{reit98} and \citet{reit00} on a field 
located $\sim 20$ kpc from M31 indicates that this trend carries 
into more distant reaches of the halo.    Whether or not [Fe/H] varies globally with radius 
is unclear, although \citet{vp92} suggested that a strong metallicity 
gradient does {\it not} exist in the inner halo based on the $(B-V)$ colors of the RGB.  

PvdB94 used star-counts to study the surface brightness profile  
of the M31 halo for $r < 20$ kpc.  Using images with sub-arcsecond seeing 
to discriminate against most of the faint background galaxies that are the primary
source of field contamination, they were able to reach an equivalent surface
brightness of $\mu_V \sim 29$, a factor of 10 fainter than reached 
by conventional surface photometry.  They concluded that the entire halo 
profile along the SE minor axis of M31 was well fit by a single 
$r^{1/4}$ law profile, falling off steeply at the outer reaches of their survey.  
However, the PvdB94 CCD fields were quite small by today's standards
(2$^{\prime}$ $\times$ 3$^{\prime}$), severely limiting the statistical weight
of the data.

In the present paper, we further the star count technique of 
PvdB94 to study the outer M31 halo, 
employing significantly larger fields with contemporary CCD mosaic cameras
to detect the sparsely spread stars expected in the outer halo.   
Our primary goals are to derive the 
stellar density profile, the extent, and the chemical composition of the outer 
M31 halo, as well as search for substructure that may be the result of 
smaller dwarf galaxies accreted by M31, since the presence of such streams has been 
claimed for the MW halo 
\citep[eg.][]{hel99, ib00, yan00}.  We present here the first 
results from our study, that of a field located $\sim 20$ kpc from the M31 center
along the SE minor axis.   The material for other and more remote fields 
will be discussed in future papers.

\section{Observations + Data Reduction}

Our first set of observations is a field at 
$\alpha_{2000} = 0^h 48^m 30^s$, $\delta_{2000} = +40^{\circ} 
17^{\prime} 54^{\prime \prime}$ (labeled $\mathcal{M}2$ in the nomenclature 
of PvdB94) located 1\degpoint 5 SE of the 
M31 nucleus and roughly along the minor axis. We used the UH8K camera 
at the Canada-France-Hawaii Telescope (CFHT) on the nights of Sept. 19--23, 1996;
three of the four nights were photometric.  Exposure 
times were 4 $\times$ 900s in $V$ and 3 $\times$ 900s in $I$, and the seeing 
on most images was 0\secpoint 6 to 0\secpoint 7 FWHM.   
The location of the $\mathcal{M}2$ field is plotted in Figure 1, which 
also illustrates the positions of fields from PvdB94.

Images with similar exposure times as the  $\mathcal{M}2$ field were 
also taken of a background field at $\alpha_{2000} = 1^h 20^m 35.5^s$, 
$\delta_{2000} = +41^{\circ} 15^{\prime} 43^{\prime \prime}$ 
located 7\degpoint 1 E of M31. This field (called  $\mathcal{R}1$) was 
chosen to be far enough away from M31 to be free of M31 stars,
yet still close enough to provide an adequate control dataset (note that 
this field is {\it not} the same as the `R1' field of PvdB94).   The seeing 
for the $\mathcal{R}1$ field was the same as for the $\mathcal{M}2$ field, 
allowing galaxy/field star discrimination to be similar for 
both fields.  The Galactic latitudes of both fields are 
$b= -23^{\circ}$ for $\mathcal{M}2$ and $b= -21^{\circ}$ for $\mathcal{R}1$, 
so we expect only small differences in foreground stellar contamination and reddening.  
Images of 3 other fields (fields $\mathcal{M}3$, $\mathcal{M}4$, and 
$\mathcal{E}2$ from PvdB94) are located further from M31 and will be discussed 
in later papers.

The UH8K camera is an 8-CCD mosaic camera (each CCD with 2048 $\times$ 
4096 pixels), with a total imaging area of 8192$^2$ pixels, or a total 
area of 28\minpoint 1 $\times$ 28\minpoint 1 (scale = 0\secpoint 206 per pixel).   
One of the chips (chip 4) showed considerable bleeding 
and other inherent substructure, and was not used in any further analysis.  
The low QE of another chip (chip 6) was also problematic for the faint photometry 
crucial to this project, and was thus also not used.  The results in this 
paper are therefore based on 3/4 of the raw UH8K array, covering a rectangular area 
of 21$^{\prime}$ x 28$^{\prime}$ on the sky.

\subsection{Data Pre-processing}

The program images were pre-processed with bias frames, dark
frames and flat-fields,  combined 
in the normal manner through IRAF\footnote{IRAF is distributed by the 
National Optical Astronomy Observatories, which are operated by the 
Association of Universities for Research in Astronomy, Inc., under 
cooperative agreement with the National Science Foundation.}.

The primary goal in developing flat-field images for this study was to
flatten {\it each individual chip} as much as possible, rather than
the entire array at once.  We extensively tested different 
flat-field strategies (described below) in order to derive the best flats.
$I$ and $V$ twilight flats were obtained, but were binned 2 $\times$ 2
in order to compensate for the large (7 min) read-out time for the
entire array.  All flats were then medianed together, resulting in a
`master' twilight flat for each chip in each filter.  Pre-processing 
with these flats alone yielded science images flat to $\sim 1\%$.

Dark-sky flats, or `super-flats' (often used for superior
flat-fielding in large-field CCDs) were also developed from our (dithered) 
M31 program fields (20 in $V$, 15 in $I$ for our total sample).   
These superflats yielded 
globally flatter images (typically $\sim 0.3\%$ in I, $\sim 0.5\%$ in V) 
than the twilight flats, but using the superflats had the small disadvantage 
of leaving small `pits` on the images which were artifacts from the 
numerous bright, saturated stars on each image.  Numerous algorithms to 
completely remove these pits had limited success.  However, the pits 
yielded at most $\sim 1\%$ local deviations from the mean sky level, 
and would have no significant effect on the photometry presented here.

The program images were flat-fielded with either the superflats 
mentioned above or a `master' flat (where the twilight flats were divided by 
a heavily smoothed superflat), whichever yielded the flatter image 
(typically $\sim 0.5\%$ peak-to-peak for each chip).

\subsection{Calibration}

The photometry was calibrated through \citet{lan92} standards observed on the 
photometric nights of the observing run.  The standards were placed within 
chip 1 of the mosaic. We created equations for each chip $n$ using average 
extinction co-efficients from \citet{lan92}, employing the method of \citet{har81} :

\begin{equation}
 V = v_n - 0.152 {\rm X} + b{\rm (V - I)} + z({V_1}) + \Delta V_{1,n} 
\end{equation}
\begin{equation}
  I = i_n - 0.061 {\rm X} + c{\rm (V - I)} + z({I_1}) + \Delta I_{1,n}
\end{equation}

\noindent where $X$ is the mean airmass,
$z(V_1)$ and $z(I_1)$ are the zeropoint values derived for chip 1
of the array, and $v_n$ and $i_n$ are the instrumental aperture 
magnitudes (aperture radius of 3.3$^{\prime\prime}$, and normalized to 
an exposure time of 1 second) on a given chip $n$.  

  As each chip of the mosaic has slightly different QE values, 
we assumed the colour terms ($b$, $c$) of all chips were the same as that for chip 1, 
and simply made corrections ($\Delta V_{1,n}$, $\Delta I_{1,n}$) 
in the zeropoint $z$ for each chip in each filter.
These zeropoint differences relative to chip 1 
were derived from the observed ratios in the sky level.  
As we had numerous images in each filter to work with, 
these zeropoint corrections were well determined, with less than 0.01 mag 
scatter in all cases.  There were also no night-night variations in these terms.  
As a test of our calibration procedure, 
we reduced images of the globular cluster M92 taken during the same observing run, 
and found no significant color or magnitude shifts (chip-to-chip) after applying 
the derived corrections.  
For all photometric nights, the color term $c$ in the $I$ equation 
was consistent with zero, and thus ignored.  The color term $b$ in the $V$ equation was $b = -0.034$.   
The rms scatter of the standard stars in each of the calibration equations was 0.018 -- 0.027 magnitudes. 

\subsection{Photometry + Image Classification}

For each of the fields $\mathcal{M}2$ and $\mathcal{R}1$, the individual CCD images were 
re-registered, scaled and combined, to construct a single combined 
image for each filter/chip/field.    Each image was 
weighted before combination so as to maximize the S/N of the final image :

\begin{equation}
w_i \propto    {{I}\over{z\ \sigma^2}}  
\end{equation}

\noindent where $I$ is the integrated intensity of a star on image $i$, $z$ is the median sky value, 
and $\sigma$ is the stellar FWHM.  This function assigns additional weight on images 
with darker sky levels and smaller stellar images.

   Photometry of the objects on each combined image was performed 
with the stand-alone versions of the DAOPHOT II / ALLSTAR packages 
\citep{stet87, stet90, stet92}.  A single pass of DAOPHOT II $+$ ALLSTAR with a 
3.5$\sigma$ detection threshold was used; since
stellar crowding is not significant, a second pass of DAOPHOT 
did not add significantly to the number of objects detected.  
A stellar point-spread-function (PSF) was derived from 5-15 bright, uncrowded 
stars per image.   A constant PSF was found to adequately fit the 
data on all images.  

For this project, we are only interested in objects on the images with 
stellar appearance, so any resolved 
background galaxies that were not already rejected by DAOPHOT II were subsequently 
removed with a combination of image parameters: the DAOPHOT $\chi$ 
parameter \citep{stet87}, the $r_{-2}$ and the $\Delta m$ image moments 
\citep{kron80}.  The first two are particularly effective 
image discriminators \citep[eg. ][]{stet87,har91, mcl95}.    The appropriate image 
classification criteria (ie. values which separated the clearly non-stellar objects from stellar ones) 
were derived for each chip for each filter, and all objects exceeding any {\it one} of these criteria were 
considered to be non-stellar and rejected from further analysis.  These same criteria were used in the 
artificial-star experiments 
that followed (see below), which we used to confirm that 
very few stellar objects were rejected by any of the adopted criteria.

\section{Color-Magnitude Diagrams}

The separate $V,I$ results from ALLSTAR and the image 
classification algorithms were merged (with a matching radius of 1 pixel) 
to create $I$, $(V-I)$ color-magnitude diagrams for each field. 
That of the $\mathcal{M}2$ field is illustrated 
in Figure 2, and the background $\mathcal{R}1$ field in Figure 3.   
The total usable area from the $\mathcal{M}2$ field is 563.6 arcmin$^2$,
while that of $\mathcal{R}1$ 
is 548.5 arcmin$^2$ (or 0.9732$\times$ that 
of $\mathcal{M}2$); some area in each chip was lost 
due to the combination of dithered images, as we used only 
those parts that contained data visible in all 
separate images to maximize S/N.  Both Figures also include error bars 
that represent typical errors in $I$ and $(V-I)$ for stars with 
$(V-I) = 1.0$, as determined through artificial star experiments (see next section).
The CMDS in Figs.~2 and 3
do not include any objects classified as nonstellar
(see previous section).  While stars from all six CCDs 
have been plotted together, the CMDs from the individual chips have
slightly different photometric completeness limits (see next section) due to 
slight QE differences.

The $\mathcal{M}2$ CMD (8691 stars) contains almost twice as many
stars as the $\mathcal{R}1$ field (4648 stars), indicating the clear presence
of the M31 halo at this distance.  
Observations of this field by PvdB94 with their far smaller field size 
yielded only a statistically marginal 
excess of M31 halo stars.  But as also found by \citet{reit98}, 
the background contamination in this field is large, making the explicit use
of a background field extremely helpful to extract the intrinsic properties
of the M31 halo population.  As is apparent from Figs.~2 and 3,
both the halo and background CMDs 
show a large number of foreground Galactic dwarfs with $17 < I < 20.5$. 
The total numbers of these stars are quite similar 
in both CMDs, confirming that we can use $\mathcal{R}1$ directly for
field subtraction.

\subsection{Artificial Star Experiments}

A prime difficulty in using CCD mosaic data is accounting for the 
slight QE differences between the chips, particularly when treating the 
combined dataset as a whole.   To ascertain both photometric 
uncertainties and photometric incompleteness in our data, we used the 
traditional method of adding stars of known brightness to the science frames 
and re-reducing them.   As we are interested in using CMD location 
(magnitude and color) to isolate likely M31 halo stars, tests were carried out 
to get statistics on stars in different parts of the ($I$, $V-I$) CMD as well.
 
For each chip of the mosaic, we added a total of 40000 stars 
(10 runs of 4000 total stars added per run) to the {\it star-subtracted} 
science images, and reduced the frames in precisely the same way as 
described above -- a single pass of DAOPHOT II/ALLSTAR, merging of 
the resulting $V$ and $I$ datasets, and removal of non-stellar images using the 
image-classification algorithm.    The number of artificial stars recovered is similar 
to that found on the original images, showing that we have adequately re-created 
the crowding conditions of the original images.

In order to quantify the photometric completeness and uncertainties over the 
part of the CMD of interest, we added stars in a 2D grid over the range 
$20.5 < I < 24.0$, $0.0 < (V-I) < 3.0$.    
The magnitudes were chosen from a steeply rising luminosity function to 
mimic the LF of the stars in the original science frames.  
This is illustrated in Figure 4, where the results 
of all {\it recovered} artificial stars from chip 1 (as an example) 
are plotted, and shows not only the photometric incompleteness towards the 
lower right part of the CMD, but also the increase 
in photometric uncertainty.   From these experiments it is clear 
that $V$ incompleteness is the limiting factor in our data.   

The limiting magnitudes ($m_{lim}$; defined as the 50\% completeness level) were 
derived from all of the artificial star data for each individual chip, 
and fitting to the following interpolation function \citep[see][]{fl95} :

\begin{equation}
 f(m) = {{1}\over{2}}\left( 1 - {{\alpha(m - m_{lim})}\over{\sqrt{1 + \alpha^2(m - m_{lim})^2}}} \right)
\end{equation}

\noindent where $m$ is either $V$ or $I$, and $\alpha$ is a parameter that measures how steeply $f(m)$ 
declines from 1.0 to 0.0; for our data $\alpha$ is typically 3.0 $-$ 4.0, the rather steep transition expected in uncrowded fields.  
The values for $I_{lim}$ and $V_{lim}$ for each chip are listed in Table 1.

Because the photometric incompleteness varies as a function of $I$ and $(V-I)$ (and on the chip $i$), 
we have mapped out the function $f(I,V-I)_i$ using linear interpolation between the grid points sampled in 
our experiments.    We have used this to assign each star in Figures 2 and 3 a completeness fraction $f$
based on $f(I,V-I)_i$ {\it for the chip 
the star was measured on}.   While the presented CMDs do not represent photometrically homogeneous 
datasets, this is accounted for by the assigned $f$ values.

\subsection{Reddening}

The reddenings $E(V-I)$ of our fields are expected to be similar to that of 
M31 itself, since it is at high enough Galactic latitude that the reddening
should not change steeply with location. The local 
HI column densities from \citet{bh84} yield $E(V-I) = 0.10$ \citep[where we adopt
$E(V-I) = 1.25 E(B-V)$;][]{car89, bar00}.  

To fine-tune the foreground reddenings a bit further,
we have used our CMDs to derive the specific values for each of
our program fields, knowing that 
most of the blue ($V-I\ \sim 0.6$) stars {\it brighter} than $I=20.5$ on our CMDs are 
foreground Milky Way halo stars.   It is clear from Figures 
2 and 3 that there is a relatively sharp 
blue `edge' to the distribution of 
these stars, and we interpret this edge as representing
the bluest (ie. most metal-poor) turnoff stars in the MW halo.  Only a very 
small fraction of these stars may be blue HB 
stars in our halo, as there are significantly more turnoff stars in an 
old stellar population.    Using the oft-observed [Fe/H] $\sim -2.2$ 
globular cluster M92 as a template for such metal-poor stars, we derive a 
turnoff colour  $(V-I)_o = 0.534 \pm 0.015$ based on the fiducial 
sequence of \citet{jb98}, 
and its well determined reddening of $E(V-I) = 0.025 \pm 0.010$ 
where we have assigned a further uncertainty of $\pm 0.01$ 
to the Johnson \& Bolte turnoff color.   This value is consistent with 
that derived from the turnoff of the similarly metal-poor cluster
M30 \citep{san99}, although the reddening for M30 is less certain.  

In our M31 fields, stars brighter than $I=21$ and redder than $(V-I) = 0.4$ were
used to define color histograms for both fields, as shown in
Figure 5.  To make an unbiased estimate of the ``edge'' color,
we first constructed a smoothed histogram from the raw data by
treating each star as a unit Gaussian with width $\sigma(V-I)$
and summing all the Gaussians.  An edge detection algorithm
(basically, a numerical second derivative of the smoothed histogram)
was then used to estimate the color at which the numbers of field
stars begin to rise most steeply.  For the $\mathcal{M}2$ field, we find
$(V-I)_e = 0.635$, and for the background field $\mathcal{R}1$, we find
$(V-I)_e = 0.615$.  Smoothing kernels in the range $\sigma (V-I) = 0.003$
to $0.02$ were tried but changed the results only at the
$\pm 0.005$ mag level. We adopt $\pm 0.01$ mag for the internal
uncertainties in each edge color.

From this analysis we obtain $E(V-I) = 0.10 \pm 0.02$ for the $\mathcal{M}2$ 
field and $E(V-I)= 0.08 \pm 0.02$ for the $\mathcal{R}1$ field, and we adopt
these values for the following discussion.
Note that the uncertainties in the absolute reddenings are $\pm 0.02$, 
but only $\pm 0.01$ for the reddening {\it difference} between the two.
Both are consistent 
with the reddening estimates of \citet{bh84} and \citet{sch98} for these locations.

\subsection{TRGB Distance to M31}

The $I$-band luminosity function of the $\mathcal{M}2$ field can be used to identify
the magnitude level of the tip of the RGB (TRGB), and thus
the distance modulus of M31.  Before doing this, it is 
advantageous to use the $\mathcal{R}1$ field to subtract
the background luminosity function from $\mathcal{M}2$.  
For the background field, we first added $\Delta I = 0.03$ 
and $\Delta(V-I) = 0.02$ to all the stars
to account for the slight reddening difference between $\mathcal{R}1$ 
and $\mathcal{M}2$ (see above).  
Then, to optimize the count statistics, we first removed
from both CMDs the bright stars ($I < 20.5$) bluer than 
$(V-I) = 1.4$ or redder than $(V-I) = 3.0$, as well as a few fainter
stars well to the blue of the RGB population.  Next, we constructed
a completeness-corrected LF for each field by representing each
star with a Gaussian of width $\sigma(I)$ and area $1/f$ where
$f$ is the completeness factor at that magnitude and color (see above),
then summing all the individual Gaussians.
Finally, the residual luminosity function of the M31 giant branch was defined
as $LF(I) = LF(\mathcal{M}2) - 1.0275~LF(\mathcal{R}1)$ (where the factor 1.0275 is the
total area of $\mathcal{M}2$ relative to $\mathcal{R}1$).  

Figure 6 shows the resulting LF, now fully corrected for background and
incompleteness.  The particular case shown is for
a smoothing kernel $\sigma(I) = 0.02$, though we tried values from
0.01 to 0.05 with no noticeable differences.  Clearly, for $I<20$, 
the ($\mathcal{M}2 - \mathcal{R}1$) subtraction cancels everything out with only
small statistical fluctuations, exactly as
it should if there is no significant population of younger, more
luminous stars sitting above the old-halo RGB tip.  Fainter than
this, we see the exponential rise of the RGB itself continuing well
past the faint limit of our data.

To estimate the TRGB, we employed the same edge-detection algorithm
described earlier to construct a numerical second derivative (slope change)
from the smoothed LF.  For all trials with different smoothing kernels,
a consistent ``edge'' or sharp increase in the slope of the LF
shows up at $I = 20.52 \pm 0.05$.    Previous estimates for $I_{TRGB}$ for M31 
halo field stars include $I=20.55\pm 0.15$ from \citet{mk86} and $I\sim 20.65$ 
from \citet{cou95}.    The M31 halo LF of \citet{hol96} also shows 
a marked increase in star counts at $I\simeq 20.6$.   Our value (which is based on a 
substantially larger sample of TRGB stars than in previous studies) is in excellent 
agreement with these values.

To determine the distance to M31, we adopt $M_{I,TRGB} = -4.1 \pm 0.1$ for 
metal-poor TRGB stars, based on Milky Way globular cluster 
RGBs and Hipparcos-based globular cluster distances \citep[see][]{har98, har99}.  The 
quoted (random) error simply represents the observed spread in the RGB tip magnitude, 
and does not include possible systematic errors on the GC distances (which are likely at 
the $\pm 0.1$ mag level).   Our adopted value 
for $M_{I, TRGB}$ is in excellent agreement 
with that derived by \citet{fer00} from 
HST-based Cepheid observations to calibrate the TRGB (and other Population II 
distance indicators).    We note that effects of halo depth along the line-of-sight 
are likely to be small ($< 0.05$ mag), as the steep radial 
gradient of the halo (PvdB94) suggests there are relatively few halo stars 
with $r_{M31} > 20$ kpc, and it is these outermost halo stars that would 
be far enough in front of M31 to create a systematic bias in the determination 
of the RGB tip (i.e., there is no strong presence of TRGB stars from 
the {\it near} side of the halo).

Using our adopted $M_{I,TRGB}$ gives $(m-M)_I = 24.62 \pm 0.11$ to M31, or a 
true distance modulus $(m-M)_o = 24.47 \pm 0.12$ once corrected for extinction 
(where $A_I = 1.48 E(V-I) = 0.15 \pm 0.03$).  
This compares extremely well with other recent estimates of the M31 
distance, including \citet{hol98} ($(m-M)_o = 24.47 \pm 0.07$ 
from M31 globular clusters), 
\citet{sg98} ($(m-M)_o = 24.47 \pm 0.04 \pm 0.05$,
 red-clump stars), and favorably with 
most earlier measurements \citep[$(m-M)_o \sim 24.3 \pm 0.1$ from a compilation
of methods discussed by ][and references within]{vdb91}.

\section{Analysis}

\subsection{Metallicity Distribution Function}

To construct a metallicity distribution function (MDF) for the M31 halo,  
we first assume that all of the stars in the CMDs are old ($t > 10$ Gyr), 
so that the location of the RGB stars in the CMD 
depends almost solely on metallicity.  We then use stars in the top two magnitudes 
of the RGB to map out the metal abundance [m/H] (defined as 
[m/H] = log($Z/Z_{\odot}$), where we have used $Z_{\odot}=0.0172$ for consistency with Harris \& Harris 
2000) defined from interpolation within a fiducial
grid of RGB model tracks.    Our technique of 
mapping the $I$, $(V-I)$ CMD to metallicity is similar to that 
employed in previous work
\citep[eg.][]{dur94, hol96, har99}. Here, for the fiducial lines we 
use the finely spaced grid of the evolutionary tracks by \citet{van00}, 
calibrated in $(V-I)$ color by standard globular cluster fiducial sequences 
\citep[such as those found in ][]{da90}.  Our methodology follows that of
\citet{hh00}, whose discussion should be referred to for more detail.

In Figures 7 and 8 we have re-plotted the $\mathcal{M}2$ and 
$\mathcal{R}1$ CMDs, overlaid 
with the evolutionary tracks for 0.8 $M_{\odot}$ stars from \citet{van00}.    
An empirical shift of $-0.03$ was applied to the $(V-I)$ color of each model in order 
to match the metallicities of observed RGBs of Milky Way globular clusters across
the entire metallicity range; 
see Figure 10 of \citet{hh00} and the discussion therein.
These models span the range [Fe/H]$=-2.31$ to $-0.40$. To calibrate the
slightly more metal-rich stars we have also added a single [Fe/H]$=+0.07$, $t=12$ Gyr 
isochrone from \citet{bert94}.   
The \citet{van00} models assume [$\alpha$/Fe]=$+0.3$ 
(a typical value for Milky Way halo stars), so that the conversion 
[m/H] $\sim$ [Fe/H] + 0.3 holds for most metallicities.  We note the 
Bertelli et al. RGB is an isochrone and not an evolutionary track, 
but the distinction is very small for stars 
in the upper RGB of an old population, and will not affect the analysis that follows.  

The model grid was then shifted by the M31 distance 
modulus ($(m-M)_I=24.62$) and the appropriate reddening as derived above.   
The method of determining [m/H] from interpolation within
the $I, (V-I)$ CMD is the same as that described by \citet{hh00}:   
briefly, $M_{bol} = M_I + (V-I)_o - BC_I$ is calculated for each star, 
where the bolometric correction $BC_I$ 
is obtained from the model tables.  Then in the ($M_{bol}, (V-I)_o$) plane bi-linear 
interpolation is performed on the pair of tracks bracketing the star, 
from which the metal abundance $Z$
(and thus [m/H]) is derived.   This method was applied to each star in the 
CMD lying in the region bounded by the model 
RGBs and in the magnitude range $20.6 < I < 22.5$.   Stars more than 
0.02 mag {\it bluer} than the most metal-poor 
track, as well as stars very near the RGB tip where bi-linear interpolation 
is less certain, were not used in the analysis.   The magnitude 
range was chosen to minimize the effects of 
photometric spread on the MDF at fainter $I$ magnitudes, and to avoid 
the large population of blue objects fainter than $I=22.5$ that are evident in the 
$\mathcal{R}1$ field (Figure 8).   These objects are most likely faint, unresolved background 
galaxies; the image classification algorithm is not able to clearly distinguish 
between stars and galaxies at the faintest levels, leaving most of them to populate 
the bottom ends of both CMDs.  Due to its strong presence in the  $\mathcal{R}1$ field, this feature 
is not an M31 stellar population, and thus not related to the AGB bump found in the M31 CMD
by \citet{gal98}, which would be located at  $I\sim 23.1$.  Even if the AGB bump is 
a strong feature in the halo CMD (which is not clear by inspection of the  $\mathcal{M}2$ CMD)
it would have no affect on our MDF analysis.

The incompleteness-corrected number of stars in each 0.1-dex metallicity bin 
for each field is listed in Table 2.   The resulting `cleaned' MDF 
(MDF(M31 halo) = MDF($\mathcal{M}2$) $-$ 1.0275 * MDF($\mathcal{R}1$)) is 
plotted in Figure 9 and listed in the final two columns of Table 2.    
The corrected count ($N_c$) for each bin is the summation of 
the $f(I,V-I)^{-1}$ values for each star in each bin, while the errors are 
simply the Poisson errors ($\pm \sqrt N$, where $N$ is the number 
of observed stars) scaled to the completeness-corrected $N_c$.  From Table 2 it 
is clear that the background contamination in this part of the CMD  
($\sim 25\%$: 3758 stars in $\mathcal{M}2$, and 958 in $\mathcal{R}1$) is 
distinctly smaller than the $\sim 55$\%
background contamination for the entire magnitude range of the data.

As a check on the effects of photometric errors and incompleteness on our MDF 
(as well as the validity of the bulk features present in the Figure 9), 
we have split the MDF into two magnitude bins ($20.6 < I < 21.6$, 
$21.6 < I < 22.5$), which are shown individually in Figure 10.   
The uncertainties in the individual [m/H] values due to the $(V-I)$ 
photometric scatter have also been 
placed above representative parts of the MDFs. The close similarity of both MDFs over 
different parts of the CMD suggests our results are not strongly 
affected by color shifts in the fainter bin, where the photometric errors are 
larger.    This also indicates that the assumption of a 
constant color term for all chips in the mosaic (see section 2.2) was acceptable, 
for more luminous RGB stars reach large $(V-I)$ colors, and extremely discrepant 
color terms would stretch or shrink the MDF compared to that derived using from
fainter stars.  In addition, the salient features of the 
MDF (the strong, sharp peak at [m/H]$\sim -0.5$ and the muted, lower-[m/H] tail) 
show up in both bins, with remarkable consistency.

We believe the metal-rich `edge' to the MDFs in Figures 9 and 10 is {\it not} due to 
photometric incompleteness in $V$, since for the stars with 
$I\sim 22$ (lower panel in Figure 10) the 
[m/H]$\sim -0.3$ ([Fe/H]=$-0.61$) model  lies well above the 50\% completeness level, 
and at this metallicity it is clear that the MDF is already falling off sharply
at the upper end.  We note also that the deep photometric studies of 
\citet{hol96} and \citet{rich96a} for inner-halo fields (which should, if anything,
be more metal-rich than ours) unequivocally 
show no evidence for a significant population of solar-metallicity stars.  

The MDF of our 20 kpc halo field clearly exhibits the moderately high
metallicity suggested by previous studies (see Introduction).  
The median [m/H] for our entire sample is [m/H] $= -0.66$ ([Fe/H]$\sim -0.96$).
\citet{reit98} reached a similar conclusion for the halo MDF 
at this distance, albeit with a much smaller field and thus a smaller 
number of halo stars.    But there is a significant metal-poor `tail' in the 
MDF, as expected from the known presence of RR Lyrae stars in
the halo \citep{pv88} and by observations of a small but significant blue component 
of the M31 HB by \citet{hol96} and \citet{rich96a}.  

To characterize the MDF further, we perform the numerical exercise of
fitting a pair of Gaussian functions to the full MDF in Figure 9. 
The resulting peak values, standard deviations
$\sigma$ and proportions in each Gaussian are listed in Table 3 and 
plotted in Figure 11.  We will refer to each peak as `metal-rich' 
and metal-poor' below, though we do not assign these two components any particular
physical reality (they overlap heavily, and a more physically based model will
be discussed below).  The errors listed in Table 3 
are the internal uncertainties in the fit.    Possible 
external uncertainties include the adopted reddening and/or distance:
A typical error in $E(V-I)$ of 
$\pm 0.02$ magnitudes produces rather small changes in all three fitted parameters : 
$\pm 0.012$ dex in the mean [m/H], $\pm 0.005$ dex for $\sigma$ and $\pm 1.6\%$ 
for the fraction of stars attributed to each peak.   
Uncertainties in the distance of $\pm 0.12$ mag have a 
larger effect, producing changes of $\pm 0.02$ dex 
for the mean [m/H], $\pm 0.012$ dex for $\sigma$, and $\pm 6\%$ for the proportions.   

It is noteworthy that the metal-poor part of the MDF (in the sense of the
two-component fit summarized above) nominally
makes up a full $40 \pm 2\% (\rm{rand.}) \pm 8\% (\rm{sys.})$ of the total 
halo population.  However, asymptotic giant branch (AGB) stars {\it bluer} than the 
RGB may have exerted small biases on this fraction
(recall that we find very few AGB stars {\it brighter} than the 
RGB tip, and any such luminous ones will have been masked out 
in our $I$ selection criteria).   The 
AGB fraction (defined here as the number of AGB stars relative to the number of RGB 
stars at the same absolute magnitude) from the evolutionary models of \citet{gir00} 
is $22 \%$ \citep[see also][]{har99}.  

How these AGB stars will affect the MDF is complex, since the color difference 
between AGB stars and RGB stars varies with both $M_I$ and metallicity.
For this reason we have chosen to look simply
at the extreme cases of AGB contribution: 
case A, where all AGB stars related to the metal-rich population stay 
within the metal-rich population (with a similar assumption for 
the metal-poor population); and case B, 
where all metal-rich AGB stars are shifted enough to the 
blue to become part of the metal-poor region of the MDF, while the metal-poor
AGB stars are shifted out of the sample altogether.
The true answer should lie between these extremes.  
In case A the proportions in each of the two components will be as derived with the 
dual-Gaussian fits described above, with no correction: $40\%$ in the 
metal-poor component, 60$\%$ in the metal-rich component.  
With case B, we find that 0.22*1674 = 368 AGB stars from the metal-richer
group will appear in the metal-poor component, so that
the total metal-poor population should be (0.402*(2800) $-$ 368)$=$ 758 stars, 
or $31\%$ of the total RGB population.  
Taking a simple average and spread of these two extremes,
we estimate that the metal-poor component 
comprises $36\pm 5\% (\rm{rand.}) \pm 8\% (\rm{sys.})$ of the total 
M31 halo population at 20 kpc.   Our results are comparable with the 
estimations of \citet{hol96} where the metal-poor component in their HST fields 
(located at 7 kpc and 11 kpc from M31) is $\sim 
25\%$ to $50\%$ of the total halo stellar population. 

\subsection{Spatial Distribution}

Across the half-degree extent of the $\mathcal{M}2$ field, there is a noticeable
gradient in stellar density which reflects the steep falloff of the M31
halo along its minor axis.  This effect allows us to
examine the spatial distribution of the metal-poor and metal-rich 
components of the M31 halo MDF defined above, and to find out if any clear
differences exist between them.  A more thorough investigation of 
the outer surface brightness profile of M31 using star-counts 
will appear in a forthcoming paper.   

For the purposes of investigating the spatial profile, we 
avoid the intermediate metallicity range where there is 
sizeable overlap between the two formally defined Gaussian components
defined above.  Thus, we define those stars with 
$-0.3 >$ [m/H] $> -0.6$ as the ``metal-rich population'', 
and the ``metal-poor population'' those with [m/H] $< -1.0$.  
For each 0.1-degree radial bin in radius from the center of M31 $r_{M31}$, 
we construct the number density ($\sigma = N_c / A$, where $N_c$ is the 
incompleteness-corrected number of stars in each annulus and $A$ is the 
area of the annulus in square arc-minutes) 
profiles for each component.  The results are plotted in log-log form in Figure 12.    
This plot shows that stars in both populations of the MDF follow a 
very steep profile, confirming the number counts seen by PvdB94.   
Fitting the data to the function $\sigma \propto r^\gamma$ with least-squares yields 
$\gamma=-5.25\pm 0.63$ (fitting uncertainty only) 
and $\gamma = -6.54 \pm 0.59$ for the metal-poor and metal-rich distributions. 
However, the $\chi^2$ values are rather large (primarily due to 
the high metal-rich point at $r_{M31}\sim 1.25 ^\circ$).     While the 
results formally indicate that the metal-richer stars are
more centrally concentrated, the statistical significance is not conclusive.
Forthcoming papers based on reductions of the rest of our dataset will 
address this question more extensively.

\section{Discussion}
 
Our derived M31 MDF in total (Figure 9) shows a strong, rather narrow 
peak at [m/H]$\sim-0.5$ (corresponding to [Fe/H]$\sim -0.8$), and an extended metal-poor tail.      
The dominant high-metallicity peak has been seen in most previous photometric studies (which surveyed 
fields located from 7 to 20 kpc from the M31 nucleus) to date, although 
the derived dispersion about the mean [Fe/H] has varied considerably.    
\citet{mk86} found [Fe/H]$\sim -0.6$, while \citet{ch91} found the halo to have a metallicity 
similar to that of 47 Tuc; both studies suggested a significant metallicity spread.     \citet{pv88} and \citet{dav93} 
both found a mean [Fe/H]$\sim -1$\footnote{The lower value of [Fe/H]$=-1$ 
from Pritchet \& van den Bergh(1988) is due to their 
use of a $V$ limit to their $V$, ($B-V$) CMD.  For the tip of 
the RGB, $M_{V}$  decreases as metallicity increases, and their chosen cutoff 
would have been biased against higher-[Fe/H] stars.}, with a smaller metallicity dispersion $\sigma \sim 0.3$ dex.   \citet{dur94} 
derived [Fe/H]$\sim -0.6$, with a similarly small $\sigma \sim 0.3$ dex.  \citet{cou95} also found a rather 
metal-rich halo ([Fe/H]$\sim -0.5$, $\sigma \sim 0.5$).  \citet{hol96} showed that the M31 halo MDF is 
asymmetric, with a peak at [Fe/H]$\sim -0.6$, but where 
stars had $-2 < $[Fe/H]$ < -0.2$.  Finally, \citet{reit98} concluded the M31 halo has [Fe/H]$< -1$ based on their 
photometry of the 20 kpc field (similar location as our $\mathcal{M}2$ field).    All studies \citep[with the exception 
of][]{dur94} based their metallicity estimates on the location of globular cluster RGBs, and in particular 
with that of 47 Tucanae, which has [Fe/H]$\sim -0.7$ \footnote{Note that in the RGB 
calibration scheme of Harris \& Harris (2000) [which is adopted in this paper] 47 Tucanae is found to have 
[m/H]$\sim -0.63$, or [Fe/H]$\sim -0.94$; which is very close to that of our metal-rich peak.   As 
most M31 halo studies have compared directly
to the 47 Tuc RGB, there is {\it strong} agreement between our results and all previous work.}.  
This observed peak in our MDF is very similar to that of other studies, 
and strengthens the conclusion that a large metallicity gradient does {\it not} exist in the M31 halo 
at $r_{M31} \sim 7 - 20$ kpc.

However, the metal-poor tail in our MDF is also strong, and our preceding analysis indicates that it
{\it cannot} be explained as AGB contamination from the metal-rich population.  
While most early studies focussed primarily on the 
mean [Fe/H] of the {\it entire} halo population and the consequent 
high dispersion $\sigma$ about this mean, 
it is clear that such a description covers over many of the details in the MDF.
We will explore how the MDF gives us some clues as to the origin of the 
M31 halo via comparisons with the globular cluster system, and 
with predictions from 
simple models of galactic chemical evolution.   We will also compare 
our results to those from two elliptical galaxies for which MDFs have been derived from 
resolved stellar populations: NGC 5128 \citep{har99, hh00} and M32 \citep{gri96}.  

\subsection{Halo or Bulge?}

The presence of a strong metal-rich population 
in the outer regions of M31 has raised the 
question whether or not we are looking at 
a true `halo' population \citep[as assumed by many works, including][and 
the current study, to name but a few] {mk86, dur94, hol96}, or an outward extension
of the rather large `bulge' of M31 \citep[][]{bica91, free96}.  

While difficult to study because of intense crowding, the M31 nuclear 
(or inner) bulge is similar to that of the MW in that it seems to be dominated by 
old stars of roughly solar metallicity \citep{renzini99,jablonka99, jablonka00}.  
Of note here is that the numerous M31 halo studies to date (including the 
present study) do {\it not} find the Solar-metallicity stars that make up
the inner bulge population.  
The mean [m/H] in the bulge populations suggests a rather strong metallicity gradient 
($\sim 0.1$ dex/kpc or more) in the M31 bulge, while the relative uniformity 
of the mean metallicity of the {\it halo} population 
at $r_{M31} = 7 - 20$ kpc suggests that such a gradient does 
not continue beyond the inner bulge \citep[and the discussion above]{pv88,rich96b}.     

However, it is known that the $r^{1/4}$ nature of the inner bulge continue out to $r=20$ kpc 
(PvdB94).  Thus it becomes difficult to simply define a clear `bulge' and `halo' in terms similar to that 
of the Milky Way.   It seems likely that 
we are actually observing a continuum of populations in the M31 {\it spheroid} -- a $Z\sim Z_{\odot}$ 
population in the central regions and somewhat more metal-poor in the outer reaches.   That the 
same distribution holds well for the entire observed range to date lends credence to the possibility that the 
formation of the metal-rich `bulge' or inner spheroid is strongly related to the formation (or the aftermath) 
of the `halo' or outer spheroid.   Studies of the MDF of stars in the region $r_{M31} = 2-7$ kpc (and 
similar radii in other large spheroids) could prove quite interesting in pinning down this metallicity gradient, 
and assist in understanding the origin of bulges/spheroids in M31 and other galaxies.

Based on this limited information, we are led to conclude that our 20-kpc field
is almost entirely `halo' rather than `bulge' (in the traditional sense), but perhaps more correctly 
called the `outer spheroid'.

\subsection{Comparison with the Globular Cluster System}

M31 possesses a large ($> 400$) population of globular clusters which has 
been the focus of many metallicity-based studies
\citep[][and references within]{huchra91}.  Recently, 
\citet{bar00} completed a large catalog of spectroscopically- and photometrically- derived metallicities of the 
M31 GCS and found its metallicity distribution to be 
roughly bimodal \citep[seen also by][]{huchra91, ashman93}, with peaks 
in the distribution at [Fe/H]$\sim -1.4$ and [Fe/H]$\sim -0.6$.    
In this sense the M31 halo resembles the bimodal GC distribution in the MW 
\citep{zinn85,har00}, with the metal-poor `halo' component and 
the more metal-rich `bulge' subsystem \citep{minn95, barbuy98, cote99, bar00}
at very similar mean metallicity values.  Both \citet{bar00} and 
\citet{huchra91} also found a {\it small} [Fe/H]-gradient in the population, with 
the tendency for the most metal-rich GCs to lie at smaller $r_{M31}$.   

Figure 13 shows a plot of our field MDF with that of all M31 
GCs with spectroscopically determined metallicity
values ($N=188$) from \citet{bar00} (we have applied [m/H] = [Fe/H] + 0.3 to
the Barmby et al.~results).  While we have used 
GCs at all radii around M31 for comparison, the lack of any strong metallicity 
gradient in the M31 GCS indicates this is an acceptable choice.     
What is seen is that {\it both} the globular clusters and field stars span 
the complete range $-2.2<$[m/H]$<-0.2$ \citep[see also][]{reit00}. 
However, the most striking feature of
Figure 13 is that the two distributions have emphatically different shapes:
far more GCs are metal-poor, while a far 
larger fraction of bulge/halo field stars are metal-rich.  From \citet{bar00}, $34\%$ of 
the GCs (of those that could be confidently assigned to one population or the other) are `bulge' (ie. metal-rich) clusters, 
while we find that $\sim 60\%$ of the field stars have a similar [m/H].  

This comparison suggests that the {\it specific frequency} $S_N$ \citep[number of
GCs per unit halo luminosity;][]{hvdb81} is about three times larger for
the metal-poor population than for the metal-rich population.
A very similar result was found by \citet{har99} for the
dual-component GCS and halo in the giant elliptical
NGC 5128.  {\it If} the GCs and halo stars in the same metallicity
range are generically related,
this comparison suggests either (a) the formation efficiency of the metal-rich clusters 
was lower, or (b) the formation efficiency of the metal-poor {\it stars} 
was lower, relative to the number of globular clusters formed.  We 
will return to this point below.   This comparison of course ignores
the possible effects of 
dynamical evolution and destruction of clusters in the inner regions 
of M31, but the differences in destructive efficiency between metal-rich and
metal-poor clusters would have to be dramatic indeed to produce the
factor-of-three difference that we now observe.

As noted earlier \citep[eg.][]{dur94, hol96} the M31 halo MDF {\it on average} is much more 
metal-enriched than the GCS \citep[by $\sim 0.3$ dex, typical of that seen in other galaxies,][]{har00}.  
The difference is now much more subtle with the expanded datasets, and is complicated 
by the possible relation between the GCs and field stars in each metallicity component.  The M31 halo 
clusters appear to have a more extended distribution than the M31 bulge clusters \citep{huchra91, bar00}, 
and there is a hint of a similar situation for the two `components' of the halo star population (see section 
4.2 above).    There may also be kinematic 
differences between the metal-rich and metal-poor GC subsystems out to 15 kpc \citep{huchra91, bar00}.   
It is plausible the GCs and stars in each metallicity subsystem are directly related, but kinematical studies of 
both the GCs {\it and} halo stars will be of the utmost importance in solving this issue -- such  
efforts are currently underway \citep{ perrett00,reit00}.

\subsection{Chemical Evolution Models}

We will now describe the MDF using simple models of galactic 
chemical evolution.  Here, we abandon the rather artificial division
between ``metal-rich'' and ``metal-poor' parts of the stellar population
that we used above for comparative purposes, and treat the MDF as a whole.
We ask here the straightforward question 
how well the Simple (closed-box) Model of chemical evolution
\citep[eg.][]{searle72, pagel75} applies to the M31 halo.
In this model, the cumulative $N(Z)$ (number of stars with abundance lower than
$Z$) follows

\begin{equation}
N(Z)\  \propto \ {{1 - e^{-(Z-Z_o)/y}}\over{1 - e^{-(Z_{now} - Z_o)/y}}}
\end{equation}

\noindent where $Z_0$ is the initial abundance of the gas in the ``box'',
$Z_{now}$ is the present-day abundance, and $y$ is the yield due to
nucleosynthesis.  The differential distribution $dN/dZ$ will show an
exponential decay with increasing $Z$.  If we allow metal-enriched gas to 
leave the system, expelled by supernova-driven winds 
\citep[the `leaky-box' model of][see also Searle \& Zinn 1978, Binney \& 
Merrifield 1998]{hartwick76}, then the true yield $y$ is reduced 
to the `effective yield' $y_{eff} = y/(1+c)$, where $c$ 
is a parameter that describes the fraction of gas mass lost from the box, 
but the exponential form of $dN/dZ$ remains the same.    

Since the Simple model is so well suited to the linear differential abundance
distribution, we choose to plot the MDF this way rather than in the
more traditional logarithmic form as in Fig.~9.
This form of the MDF is plotted in Figure 14.   
We have assumed $Z_{\odot} = 0.0172$ and [$\alpha$/Fe]$=+0.3$ throughout,
though these assumptions will have 
little bearing on the discussion to follow.   

In Figure 14 we have plotted the expected curves for Simple models
with $Z_0$ and various choices of $y_{eff}$, chosen to illustrate what the model
would predict over the observed range in $Z$.
Three features of this comparison immediately emerge:

\noindent (a) The model, in broad terms, matches the 
M31 data for the entire run $0.0 < Z/Z_{\odot} < 0.8$.
Notably, the number of observed stars at
the lowest abundances continue to rise towards $Z \rightarrow 0$ in the roughly
exponential form required by the model.  In some other galaxies (see especially
Harris \& Harris 2000 for the case of NGC 5128)
the lowest-metallicity stars are conspicuous by their
near-absence, and the differential distribution $dN/dZ$ actually declines
near $Z \rightarrow 0$, deviating strongly from the exponential-decay
condition.

\noindent (b) The best-fit effective yield for the M31 halo is 5 to 6 times
larger than for the Milky Way halo, for which 
best-fit values are in the range $y_{eff} \sim 0.0009$ \citep{rn91}.
If the intrinsic nucleosynthetic yield $y$ for the two galaxies was at all
similar in their earliest star-forming stages, then (within the context of
the same chemical enrichment models) the M31 halo must have successfully
held on to a much higher fraction of its gas, allowing the enrichment to 
proceed up to higher levels.

\noindent (c) There is an excess of {\it intermediate-metallicity} stars 
with $Z \sim 0.3 - 0.5 Z_{\odot}$.   Comparisons of the cumulative $N(Z)$
distributions with the enrichment models 
through a standard Kolmogorov test show that the data differ from
the models with a significance higher than 99\%, regardless of
choice of $y_{eff}$.  That is, the Simple model
matches the data to first order, but fails to describe the MDF in detail.

It would have been plausible to expect that the
most natural comparison for the M31 MDF is 
the halo of the Milky Way, which is a basically similar galaxy type.
Instead, the M31 halo even in its outer regions clearly does not resemble
the Milky Way at all, but rather
the MDFs of elliptical galaxies such as NGC 5128 \citep{hh00}
or M32 \citep{gri96}.  In Figure 15, we plot
our M31 MDF with the combined MDFs for the outer halo of 
NGC 5128 ($r \sim 20$ to 30
kpc, closely comparable to our M31 field)
derived by \citet{hh00} with the same RGB models and interpolation technique. 
Both MDFs show a dominant metal-rich peak ([m/H]$\sim -0.5$ for M31, 
$\sim -0.4$ for NGC 5128) with a long metal-poor tail. 
In detail, we see that the NGC 5128 MDF extends to slightly higher metallicity 
at the top end, and has noticeably fewer stars at the low end.
Harris \& Harris (2000) suggest that a two-phase chemical evolution model
is needed to explain the NGC 5128 MDF:  an early ``accreting box'' stage
during which unenriched gas continued to flow in to the protogalaxy while
it was going through its first rounds of star formation, until the overall
abundance had built up to $Z \simeq 0.25 Z_{\odot}$; and then a second
stage after the gas infall died away and the system approached a
standard closed-box evolution.  During this latter stage, the effective
yield for the outer NGC 5128 halo is
$y \simeq 0.006$, quite similar to what we see in M31. 
Thus for M31, invoking this first phase of combined star formation
{\it with gas accretion} at very early times seems not to be necessary.

Our M31 MDF is also 
quite similar in form to the MDF of the low-luminosity compact 
elliptical galaxy M32 \citep{gri96}.  Just as for NGC 5128, M32 
has extremely few low-metallicity stars.  We note, however, that the M32 MDF
was drawn from a sample
close to the nuclear regions, while the NGC 5128 and M31 MDFs are 
from much further out in the halo;
\citep[see the comments by][on the possibility of lower-metallicity stars lying 
at larger $r$ from M32]{gri96}.  In sum, however, the material from these three galaxies 
is rather suggestive of a common formation mechanism for massive
spheroids (elliptical galaxies and large bulges), despite 
differences in the details.

\subsection{Formation of the M31 Halo}

Hierarchical models of galaxy formation provide a plausible
physical basis for how galaxies form
\citep[eg. ][]{cole94,kauf96,klyp99,cote00a}. In this context,
we expect the metal-poor M31 halo 
to have formed from the agglomeration of protogalactic, dwarf-sized
clouds \citep{sz78,hp94}, each of which had likely formed 
some proportion of stars and globular clusters before they began 
amalgamating into the M31 halo.   Furthermore, the observations
thus far are consistent with the M31 halo and bulge (spheroid) being composed
{\it predominantly of old stars} and that the bulk of star formation
took place at very early times, during the hierarchical assembly.  
The models of 
\citet{cote00a} show that such a mechanism does an excellent 
job of explaining the metallicity distributions of both the halo 
globular clusters and halo field stars in the 
Milky Way, provided that the luminosity function of the incoming
fragments is $dn/dL \propto L^{-2}$ and there is no dissipation of the gas.  
To explain the more metal-rich M31 halo, they suggest a slightly
flatter mass spectrum $\sim L^{-1.8}$ for the fragments  
to allow more massive fragments (with their decidedly more metal-rich 
stars) to contribute more to the halo.    However, enough
metal-poor stars are still formed to match the numbers that we
observe, so that it does
not seem necessary to invoke significant additional accretion of primarily
stellar material (dwarf ellipticals) at later times.   

On the other hand, it may be necessary to invoke a special explanation
for the stars at $Z \simeq 0.3 - 0.5 Z_{\odot}$, i.e. the ones
forming the peak of the MDF in Fig.~13.  Their excess of numbers
over the standard chemical evolution models is highly significant.
These stars are far too metal-rich to have been acquired from (e.g.)
dwarf elliptical galaxies or any low-mass fragments with small
potential wells, and their presence
may suggest that a particularly large ``fragment'' or partially
evolved massive satellite was acquired early on.

Another puzzle is the contrast in globular cluster specific frequency
between the metal-poor and metal-rich ends of the MDF.
The answer may lie in when and where each type of globular cluster formed.
As suggested by \citet{har99} for the clearly similar case of NGC 5128, 
one possibility for understanding the
much higher $S_N$ for metal-poor clusters is to suppose that 
the gaseous protoclouds
merged together just {\it after} most of them had formed (metal-poor) 
globular clusters, but {\it before}
much of their local star formation had taken place. That is, globular
cluster formation (logically, in the very densest parts of the protoclouds)
would occur earliest of all, and the supernova-driven winds
might have truncated star formation in
the rest of their local parent clouds.  Later on,
the bulk of the gas -- now mostly merged
into the M31 protohalo -- would then have continued to form progressively
more metal-rich stars in the considerably larger, deeper potential well of M31,
along with the majority of metal-rich globular clusters.

If the basic hierarchical merger model is correct, then the $S_N$ pattern
we have noted here should be a common property of giant galaxies.
These speculations should be tested with additional data from other such
galaxies.
Furthermore, as mentioned above, kinematics of both halo stars and 
the globular clusters are crucial to understanding 
both the M31 bulge and halo.   If the above scenario is correct, 
we would expect the metal-rich stars {\it and} 
GCs to have rather similar kinematics.  If the collapse of the 
gas expelled by the halo stars was dissipative, 
the metal-rich stellar populations should be rotating, but slower 
than stars in the inner (nuclear) bulge.  
As for the metal-poor halo component of M31, the true rotation and velocity dispersion will depend (somewhat) on the 
contributions from later accretion, but should be noticeably less than that of the metal-rich component.   As noted, 
this appears to be the case for the globular cluster system \citep{huchra91, bar00}.

The question still remains why the M31 halo is 
{\it on average} so much more metal-rich than that of the Milky Way halo, even 
though the MDF shapes are both approximately described by Simple models of 
chemical evolution.  As other authors have discussed (cf. the references
cited earlier), the Milky Way halo distribution is suggestive of the idea
that almost all its halo stars were formed within dwarf-sized units
with small potential wells, which could sustain only very low effective
yields, and that these small systems amalgamated later with relatively
little subsequent star formation.  In M31, it seems much more necessary
to propose that {\it a very large fraction of the star formation took
place only after the gaseous fragments had merged}, so that the gas
could remain within the much larger potential well of the entire
giant galaxy, and drive upward to higher metallicity.  In this sense,
the formation of the M31 halo appears to resemble that of a giant
elliptical.

\section{Summary}

As part of our survey of the M31 halo,
we have presented $(I,V-I)$ photometry from a field located 20 kpc from the
galaxy center along the SE minor axis and also a remote background field.
Using image classification and comparison with the background field,
we have been able 
to isolate a sample of several thousand M31 halo stars, from which we 
derive a TRGB-based distance modulus $(m-M)_0 = 24.47 \pm 0.12$ 
($d = 783\pm 43$ kpc).  There are no statistically significant numbers
of M31 halo stars brighter than the classic old-halo RGB tip, consistent with
the interpretation that we are looking at a homogeneous, old stellar
population.  Reddening estimates for our fields are
$E(V-I)= 0.10 \pm 0.02$ for $\mathcal{M}2$ and 
$0.08 \pm 0.02$ for the background$\mathcal{R}1$ field, found 
from the colors of foreground Milky Way halo stars in the color-magnitude
diagrams.

Using a dense grid of RGB stellar models, we derive the metallicity
distribution function for the outer M31 halo.
The MDF contains stars of a wide range of [m/H], but with a distinct
peak at [m/H]$\sim -0.54$. 
There is no evidence for a significant population of stars with 
[m/H]$> -0.3$ at this location of the M31 bulge/halo.  
A single ``Simple model'' of chemical evolution 
does match the MDF to first order, but with some clear second-order
deviations.  The effective yield best matching the MDF ($y_{eff} \simeq
0.0055$) is about 6 times higher than that for the Milky Way halo; 
instead, it more closely resembles the MDFs for the halos of 
the elliptical galaxies NGC 5128 and M32.

We outline a scenario for the formation of the M31 halo based on
contemporary models for hierarchical merging of protogalactic gas clouds:

-- The earliest star formation took place within these dwarf-sized gas
clouds clearly before they merged, during which time the metal-poor
globular clusters and some metal-poor stars formed.

-- Agglomeration of these fragments into the global halo of M31
occurred while the fragments were primarily gaseous, 
The (mainly) gaseous proto-halo then proceeded to form 
most of its stars (as well as most of the metal-rich 
globular clusters).  Most of the gas was held within the deep, massive
potential well of the giant galaxy, allowing the gas to evolve up to
high metallicity roughly according to the closed-box chemical evolution
model.  

-- Some fraction of the now-enriched halo gas may have found its way
inward to help form the still more
metal-rich population in the inner bulge ($r < $ a few kpc).   
Although the timescale of this stage is not constrained by our data,
it is expected to be not long after the halo formation itself, 
resulting in the formation of old, solar metallicity stars 
that populate the inner bulge.

\acknowledgments

This project has benefited greatly from fruitful discussions with Tim Davidge, Pat C\^{o}t\'{e}, Steve Holland, 
Raja Guhathakurta, Steve Majewski, Harvey Richer, Greg Fahlman, St\'{e}phane Courteau, Peter 
Bergbusch, Don VandenBerg, and David Hartwick.  This 
research was supported by NSERC research grants to WEH and CJP, and PRD gratefully 
acknowledges financial support from Robin Ciardullo, Harvey Richer, and Greg Fahlman.  We also 
acknowledge use 
of facilities made available by the Canadian Astronomy Data Centre, which is operated by the Herzberg 
Institute of Astrophysics, National Research Council of Canada.

%***************************************************************************
%*************************REFERENCES*************************************
%***************************************************************************

\clearpage
%***************************************************************************
%************************* TABLES *****************************************
%***************************************************************************

\begin{deluxetable}{cccccc}
\tablenum{1}
\tablewidth{0pt}
\tablecaption{Limiting Magnitudes}
\tablehead{
\colhead{} & \multicolumn{2}{c}{$\mathcal{M}2$} & \colhead{} & \multicolumn{2}{c}{$\mathcal{R}1$} \\
\cline{2-3} \cline{5-6} \\
\colhead{CCD \#} & \colhead{$V_{lim}$} & \colhead{$I_{lim}$}& \colhead{} & \colhead{$V_{lim}$} & \colhead{$I_{lim}$}}
\startdata
0 &  24.12 & 23.90 && 24.86 & 23.74\\
1 &  24.31 & 24.03 && 24.96 & 23.88\\
2 &  24.11 & 23.67 && 24.77 & 23.47\\
3 &  24.36 & 23.95 && 24.94 & 23.72\\
5 &  24.37 & 23.96 && 24.93 & 23.79\\
7 &  23.94 & 23.77 && 24.60 & 23.53\\
\enddata
\end{deluxetable}

\clearpage

\begin{deluxetable}{rcrrcrrcrr}
\tablenum{2}
\tablewidth{0pt}
\tablecaption{Metallicity Distribution Function}
\tablehead{
\colhead{} & \colhead{}& \multicolumn{2}{c}{$\mathcal{M}2$} & \colhead{} & \multicolumn{2}{c}{$\mathcal{R}1$\tablenotemark{a}} & 
\colhead{} &\multicolumn{2}{c}{$\mathcal{M}2 - \mathcal{R}1$}\\
\cline{3-4} \cline{6-7} \cline{9-10} \\
\colhead{[m/H]} & \colhead{} &\colhead{$N_c$} & \colhead{$\sigma$}& \colhead{} &\colhead{$N_c$} & 
\colhead{$\sigma$} & \colhead{} & \colhead{$N_c$}   & \colhead{$\sigma$} \\  }
\startdata
$-2.30$ &&      4.8  &   2.4 &&     0.0 &    0.0  &&    4.8  &   2.4 \\
$-2.20$ &&    14.7  &  4.1 &&     1.1  &   1.1  &&   13.7  &  4.2 \\
$-2.10$&&     26.6  &   5.4 &&     7.5 &    2.8  &&   19.1 &    6.1 \\
$-2.00$ &&    32.1  &   6.0  &&    8.5 &    3.0  &&   23.6 &    6.7 \\
$-1.90$ &&    56.6  &   7.8  &&    6.4 &    2.6  &&   50.3 &    8.2 \\
$-1.80$&&     44.8  &   7.0 &&     4.2 &    2.1  &&   40.5 &    7.3 \\
$-1.70$&&     55.8  &   7.7  &&    3.2 &    1.8  &&   52.7 &    8.0 \\
$-1.60$&&     51.5  &   7.5  &&    7.5 &    2.8  &&   44.0 &    8.0 \\
$-1.50$ &&    82.8  &   9.4  &&   12.8 &    3.7 &&    70.0 &   10.1 \\
$-1.40$ &&    82.1  &   9.4   &&  13.8 &    3.8  &&   68.3 &   10.2 \\
$-1.30$ &&    92.6  &  10.0  &&   11.7 &    3.5 &&    80.9 &   10.6 \\
$-1.20$ &&    99.3  &  10.4  &&   14.0 &    3.9 &&    85.3 &   11.0 \\
$-1.10$ &&   155.9  &  12.9 &&    17.1 &    4.3  &&  138.8 &   13.6 \\
$-1.00$ &&   138.4  &  12.1  &&   20.3 &    4.7  &&  118.1 &   13.0 \\
$-0.90$ &&   213.7  &  15.1  &&   24.5 &    5.1  &&  189.1 &   15.9 \\
$-0.80$ &&   231.8  &  15.8 &&    29.9 &    5.7  &&  201.9 &   16.8 \\
$-0.70$  &&  272.8  &  17.4  &&   40.7 &    6.6  &&  232.2 &   18.6 \\
$-0.60$  &&  365.8  &  21.0 &&    69.5 &    8.6  &&  296.3 &   22.7 \\
$-0.50$ &&   455.8  &  25.9  &&   81.2 &    9.4  &&  374.5 &   27.5 \\
$-0.40$   && 421.1  &  24.6 &&   104.8 &   10.7 &&   316.3 &   26.9 \\
$-0.30$ &&   351.7  &  22.6  &&  144.1 &   12.5&&    207.6 &   25.9  \\
$-0.20$ &&   299.2  &  22.9  &&  166.9 &   13.7  &&  132.3 &   26.7 \\
$-0.10$  &&  208.1  &  23.0  &&  167.8 &   14.3   &&  40.3 &   27.1 \\
\enddata
\tablenotetext{a}{counts corrected to area of $\mathcal{M}2$ field}
\end{deluxetable}

\clearpage

\begin{deluxetable}{crr}
\tablenum{3}
\tablewidth{0pt}
\tablecaption{Dual-Gaussian fit to MDF}
\tablehead{\colhead{ } & \colhead{metal-rich peak} & \colhead{metal-poor peak}}
\startdata
${\rm{[m/H]}}$ & $-0.520 \pm 0.008$ & $-1.195\pm 0.055$ \\
$\sigma$ & $0.203 \pm 0.009$ & $0.450\pm 0.027$ \\
proportions & $0.598 \pm 0.037$ & $0.402\pm 0.037$ \\
\enddata
\end{deluxetable}

\clearpage
%***************************************************************************
%************************* FIGURES ****************************************
%***************************************************************************

\includegraphics[width=6.5truein]{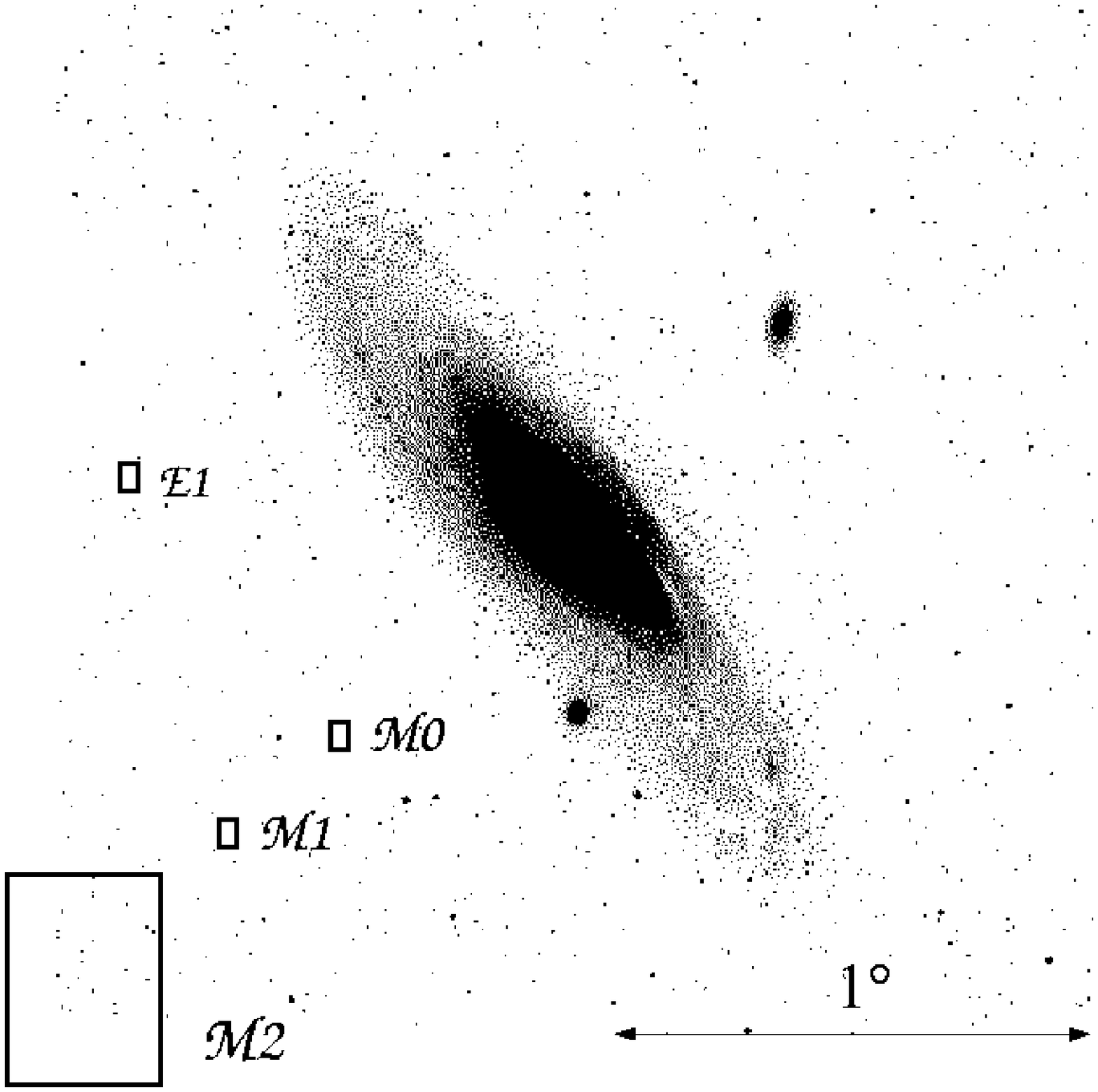}
\figcaption[fig1.ps]{Location of the M31 halo field ($\mathcal{M}$2) under study in this paper -- the field size 
is that of the usable area of the UH8K mosaic (21$^\prime$ x 28$^\prime$).  The smaller 2$^\prime$ x 3$^\prime$ 
fields represent the fields studied previously by Pritchet and van 
den Bergh (1994).  M31 image courtesy Bill Schoening, Vanessa Harvey 
/REU program/AURA/NOAO/NSF. \label{fig1}}

\includegraphics[width=6.5truein]{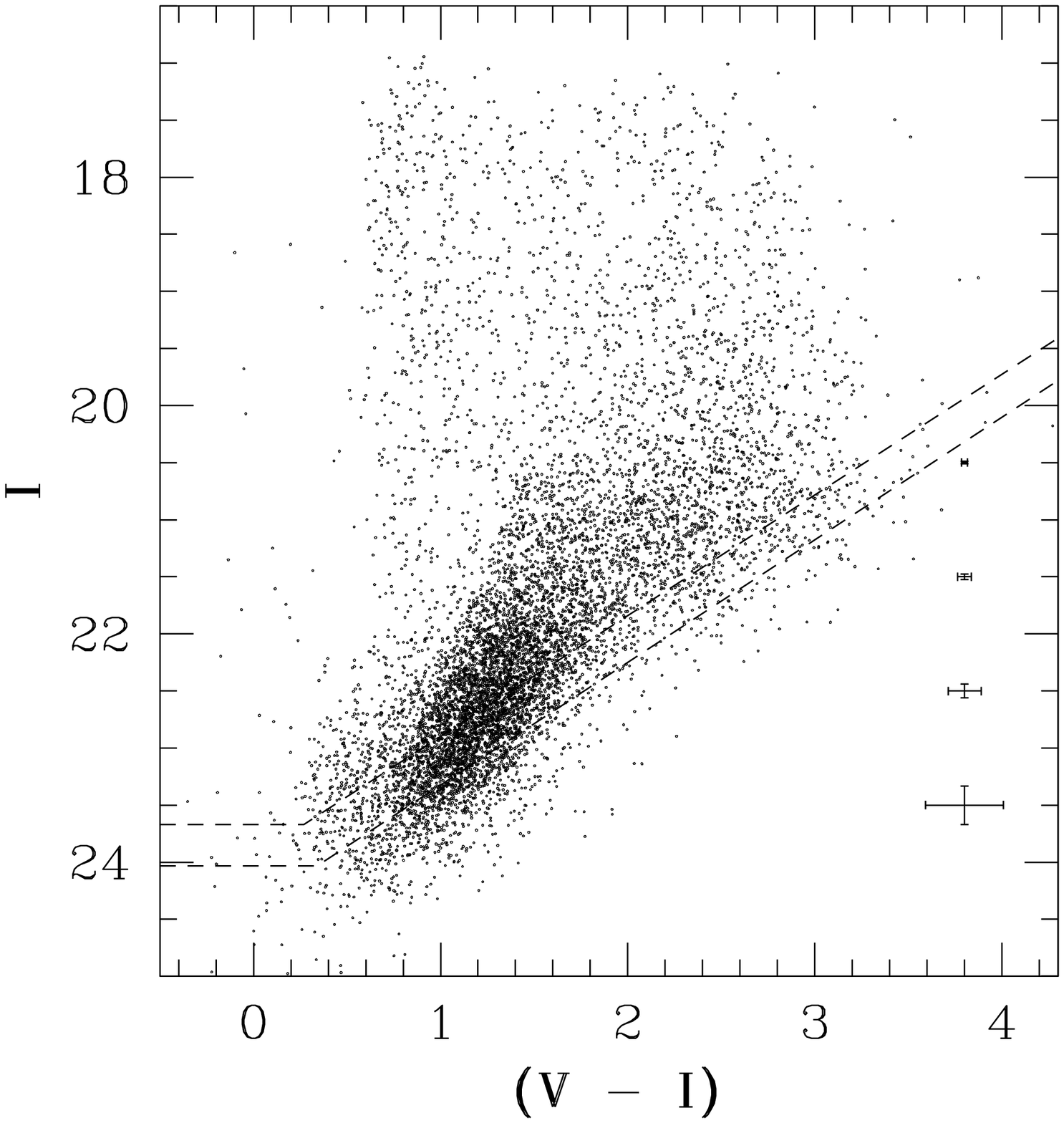}
\figcaption[fig2.eps]{$VI$ Color magnitude diagram of the $\mathcal{M}$2 halo field, based 
on data from all 6 usable chips of the CCD array.   The dashed lines denote the full range of the 
50$\%$ completeness levels for the CCDs used.    All non-stellar objects have been rejected 
via image classification.  The plotted error bars denote the {\it representative} errors for objects 
with $(V-I)= 1$.  \label{fig2}}

\includegraphics[width=6.5truein]{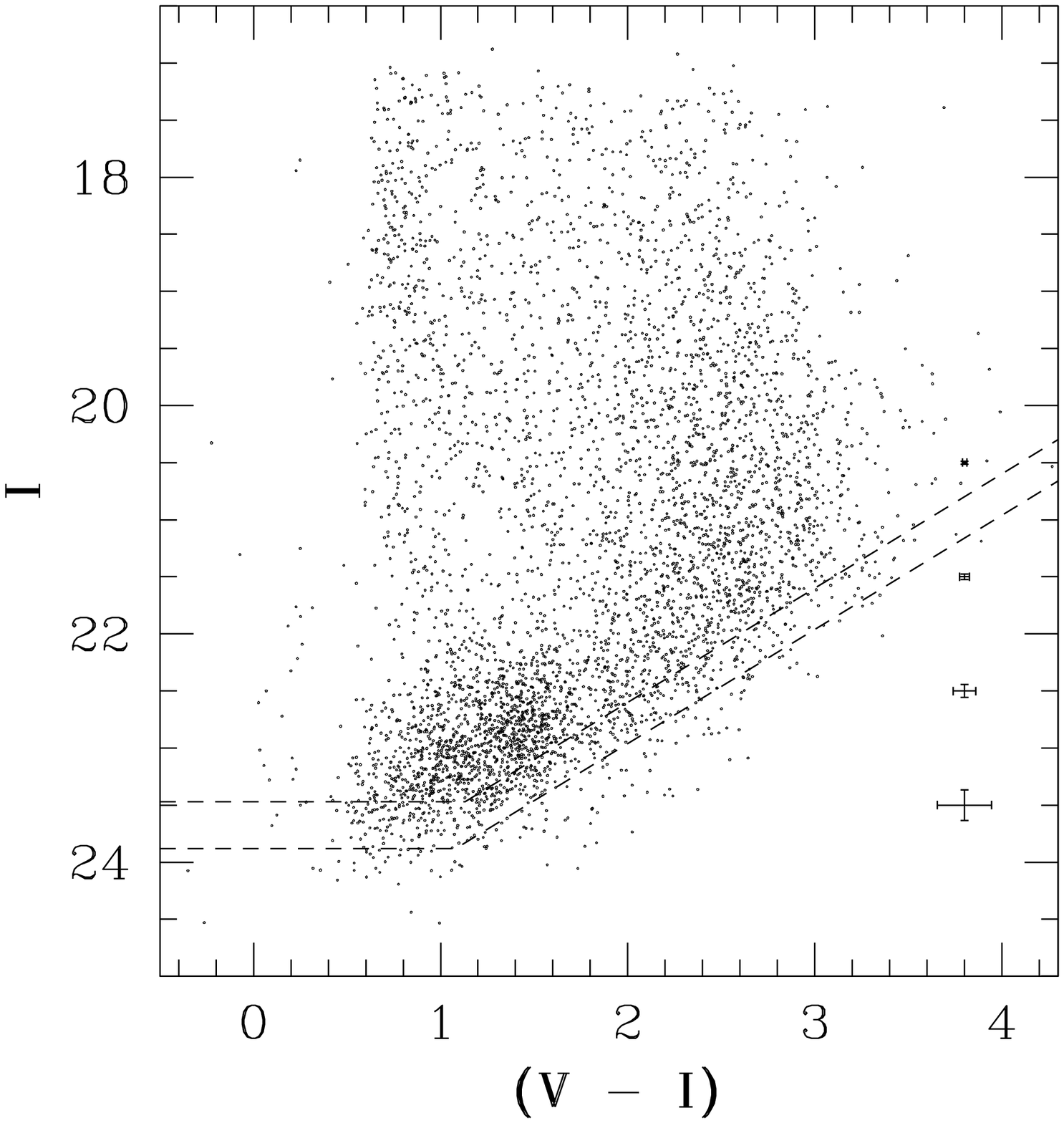}
\figcaption[fig3.eps]{Color-magnitude diagram of the background $\mathcal{R}$1 field.  Lines 
are same as those plotted in Figure 2. \label{fig3}}

\includegraphics[width=6.5truein]{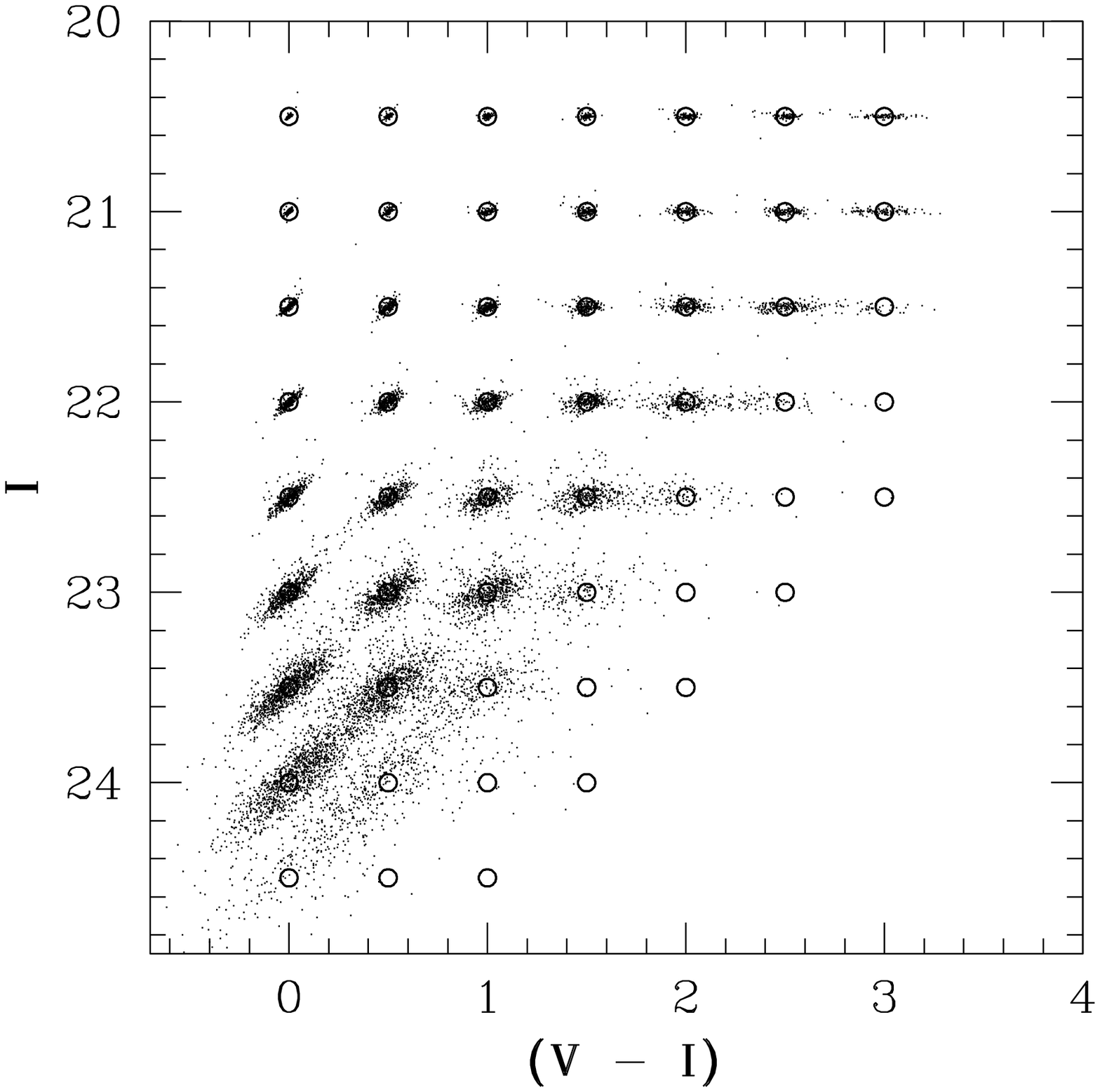}
\figcaption[fig4.eps]{Color-magnitude diagram of artificial stars recovered from simulations on chip 1
on the $\mathcal{M}2$ field (for representation purposes -- other chips are similar).  The circles show the 
discrete $I$, $(V-I)$ locations for the added stars, and the points show the recovered values.   Note the 
rather sharp transition in photometric completeness for stars with input $V\sim 24$ \label{fig4}}

\includegraphics[width=6.5truein]{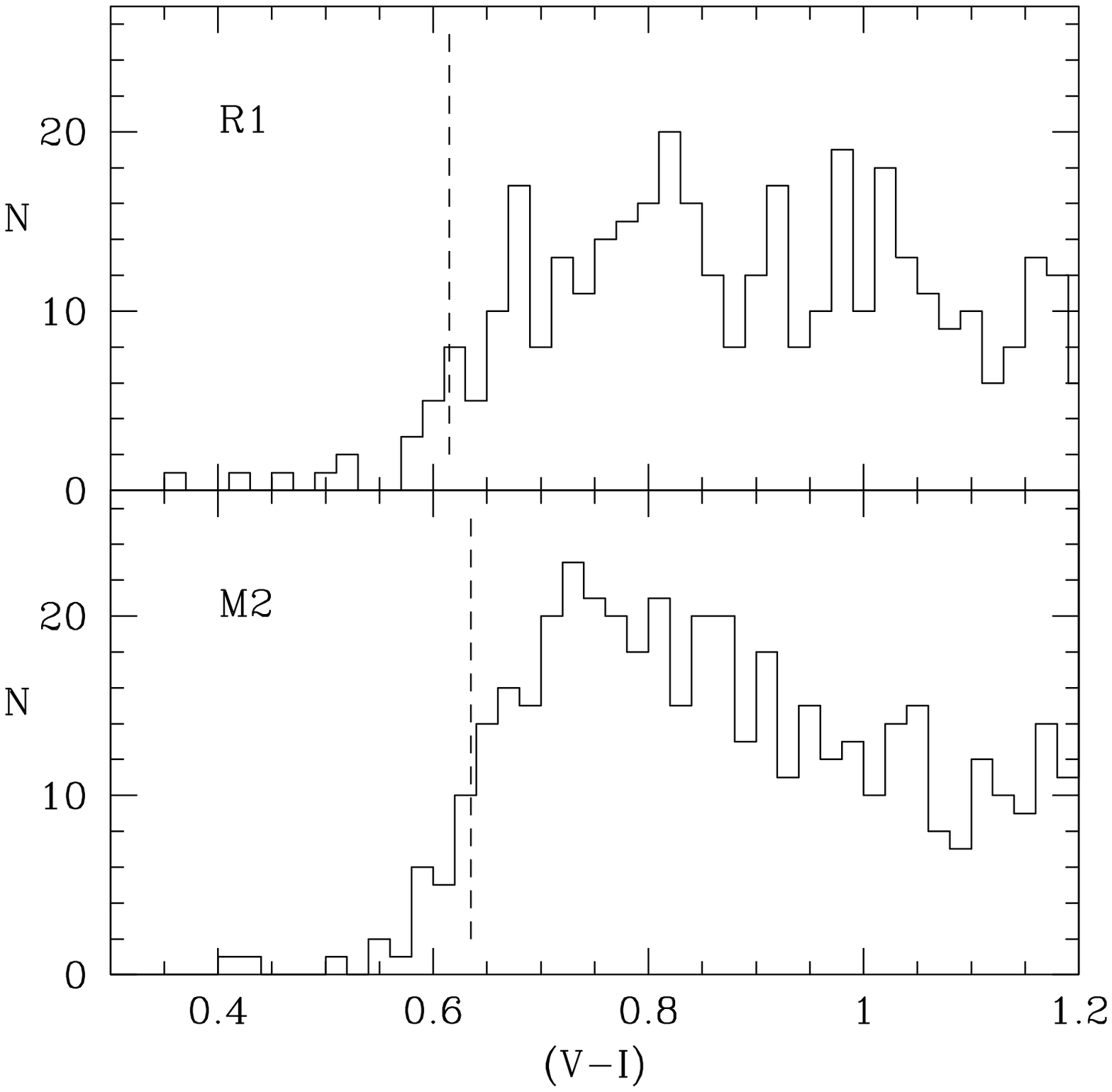}
\figcaption[fig5.eps]{Color histograms for the bright stars ($17.0 < I < 20.5$) in the 
background $\mathcal{R}1$ field (top) and the $\mathcal{M}2$ field (bottom).   In each 
case the dashed line shows the location of the blue edge of the color distribution as 
determined via an edge-detection filter (see text for details).  \label{fig5}}

\includegraphics[width=6.5truein]{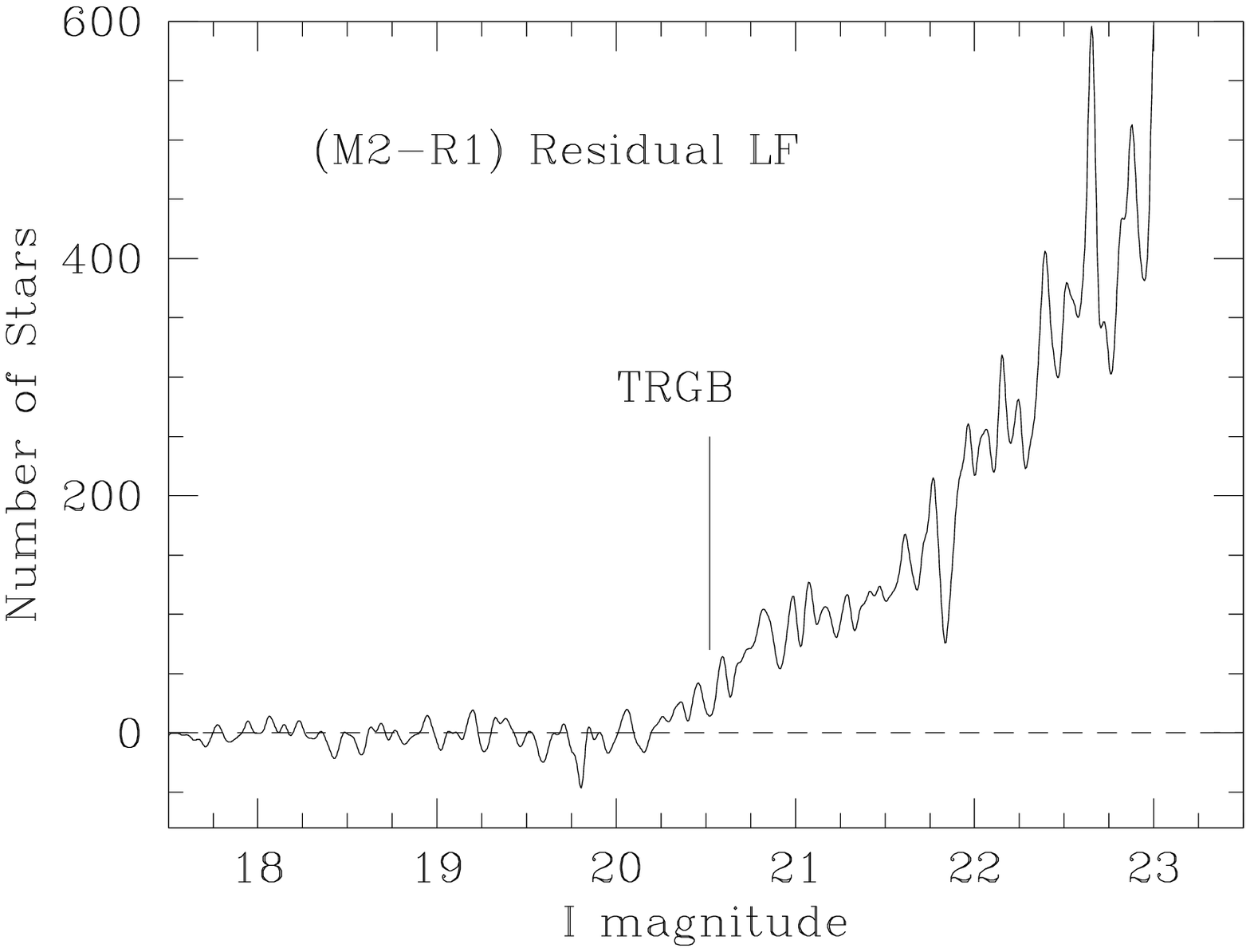}
\figcaption[fig6.eps]{Smoothed residual $I$ LF of the M31 halo field $-$ the background LF.  
The solid line denotes the position of the tip of the RGB ($I_{TRGB} = 20.52 \pm 0.05$), derived using an 
edge-detection filter. \label{fig6}}

\includegraphics[width=6.5truein]{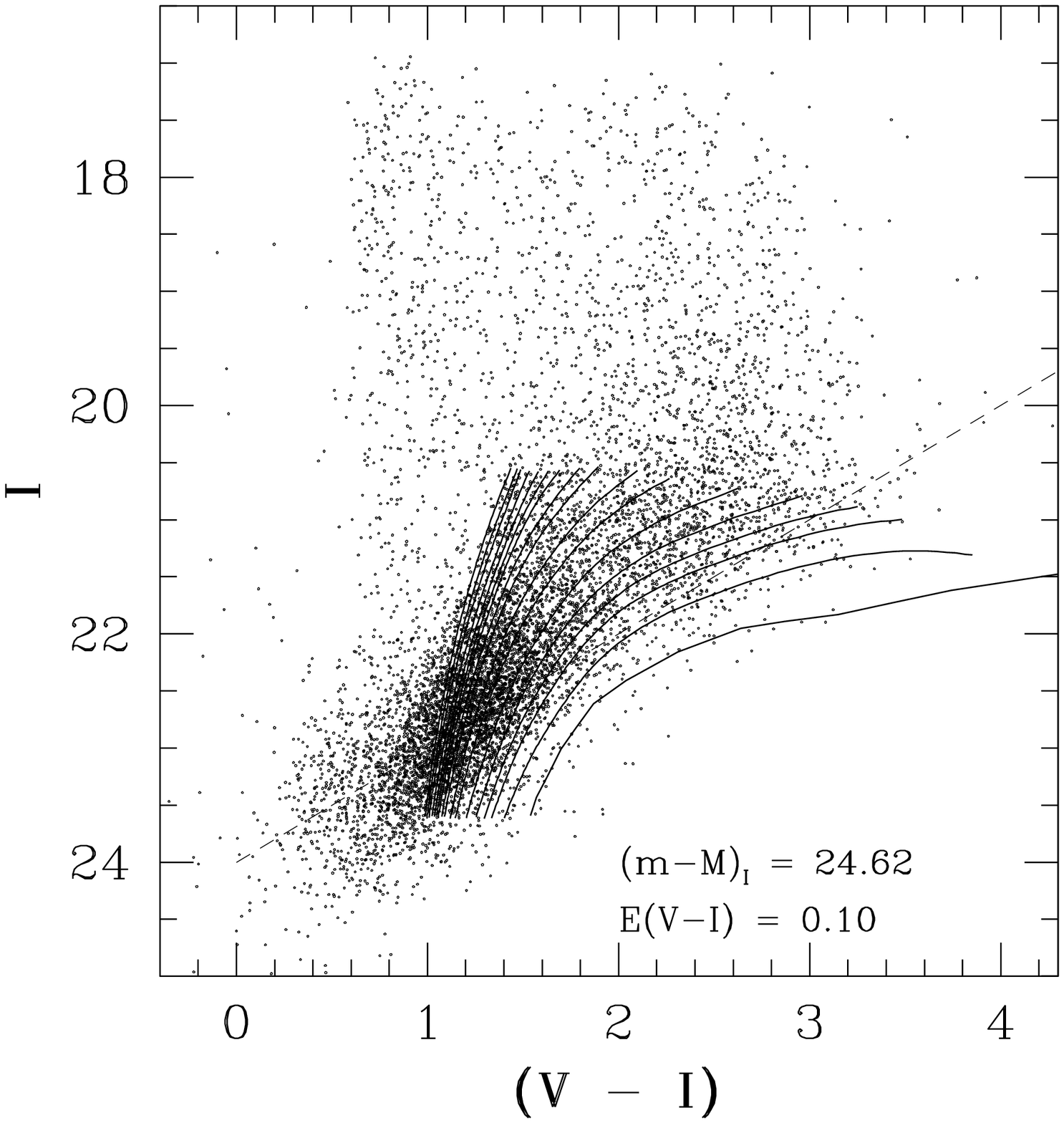}
\figcaption[fig7.eps]{CMD for stars in the $\mathcal{M}2$ field.  The solid lines are evolutionary 
tracks for 0.8 $M_{\odot}$ stars from VandenBerg et al. (2000), shifted to the given distance modulus and 
reddening.  The models have been further shifted 0.03 mag to the blue (empirical correction - see text for more 
details).   From left to right : [Fe/H] = $-2.31$, $-2.14$, $-2.01$, $-1.84$, $-1.71$, $-1.61$, $-1.54$, $-1.41$, $-1.31$, $-1.14$, 
$-1.01$, $-0.83$, $-0.71$, $-0.61$, $-0.53$ and $-0.40$.   The rightmost model is the [Fe/H]$=+0.07$ 
isochrone from Bertelli et al. (1994).  The dashed line represents the 50$\%$ completeness level for the 
least-sensitive chip in the mosaic.  \label{fig7}}

\includegraphics[width=6.5truein]{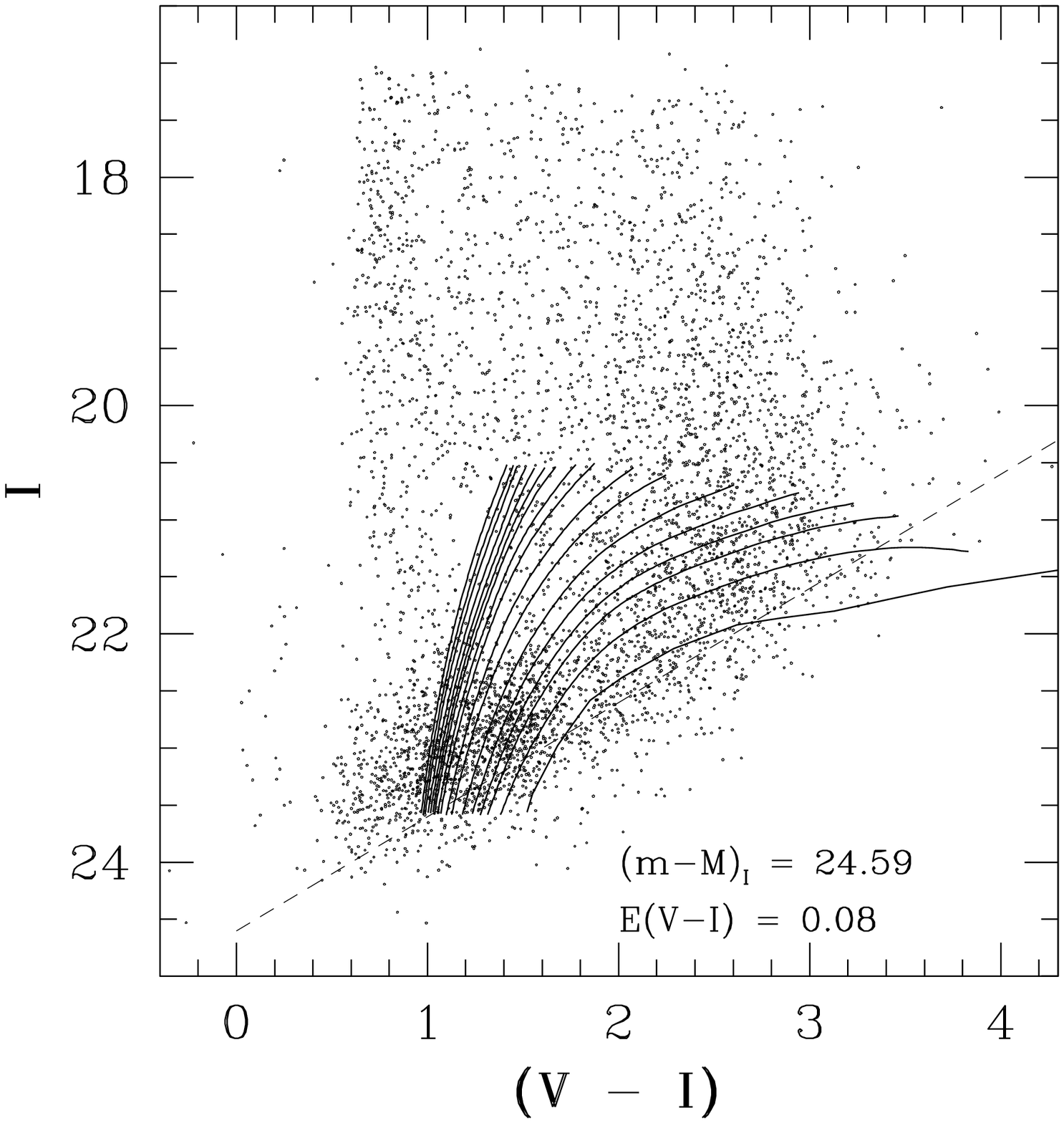}
\figcaption[fig8.eps]{CMD for stars in the $\mathcal{R}1$ field.  Solid lines are the 
same as in Figure 7.  The dashed line represents the 50$\%$ completeness level for the least 
sensitive chip in the mosaic.  \label{fig8}}

\includegraphics[width=6.5truein]{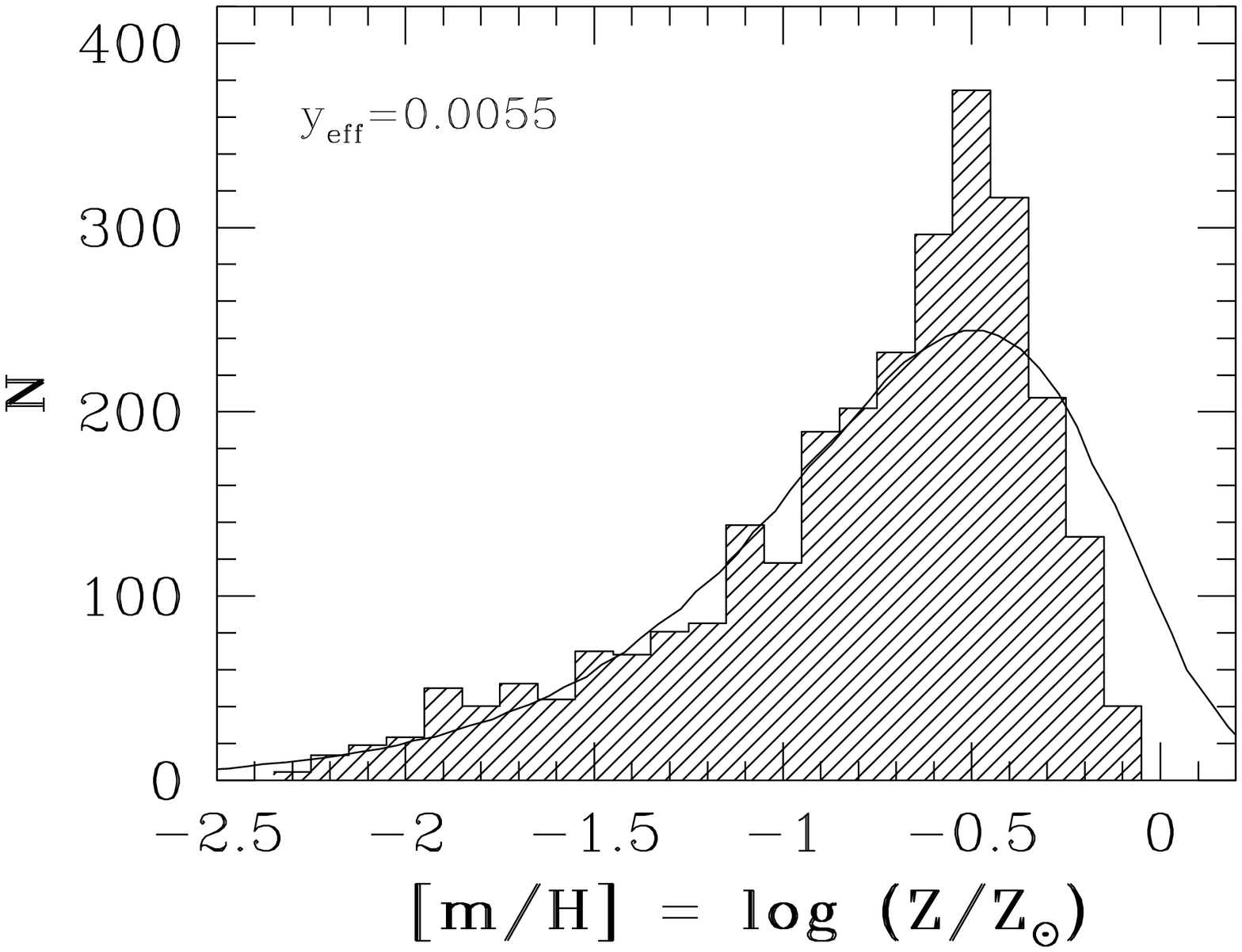}
\figcaption[fig9.eps]{Metallicity Distribution Function (MDF) for all stars 
with $20.6 < I < 22.5$.  The data has been corrected for both photometric completeness 
and for background contamination; see text for details.     The solid line denotes a single-zone, 
closed-box chemical evolution model with a yield $y=0.0055$.
\label{fig9}}

\includegraphics[width=6.5truein]{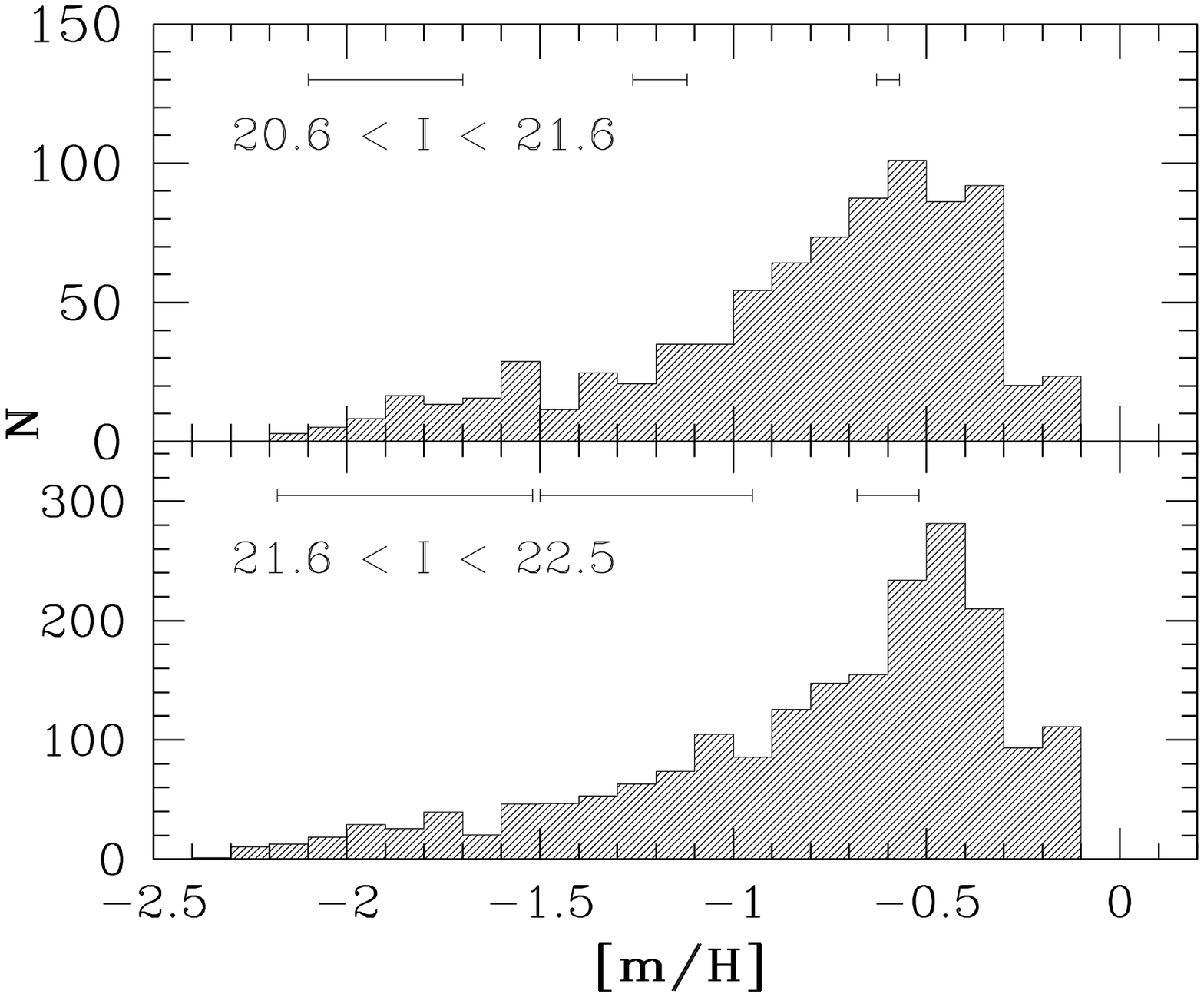}
\figcaption[fig10.eps]{MDFs for M31 halo stars in the given magnitude ranges.  
The error bars represent the uncertainties in [m/H] due to photometric errors at $I=21$ (top) and 
$I=22$ (bottom).   As in Figure 9, all data has been background and incompleteness corrected.  \label{fig10}}

\includegraphics[width=6.5truein]{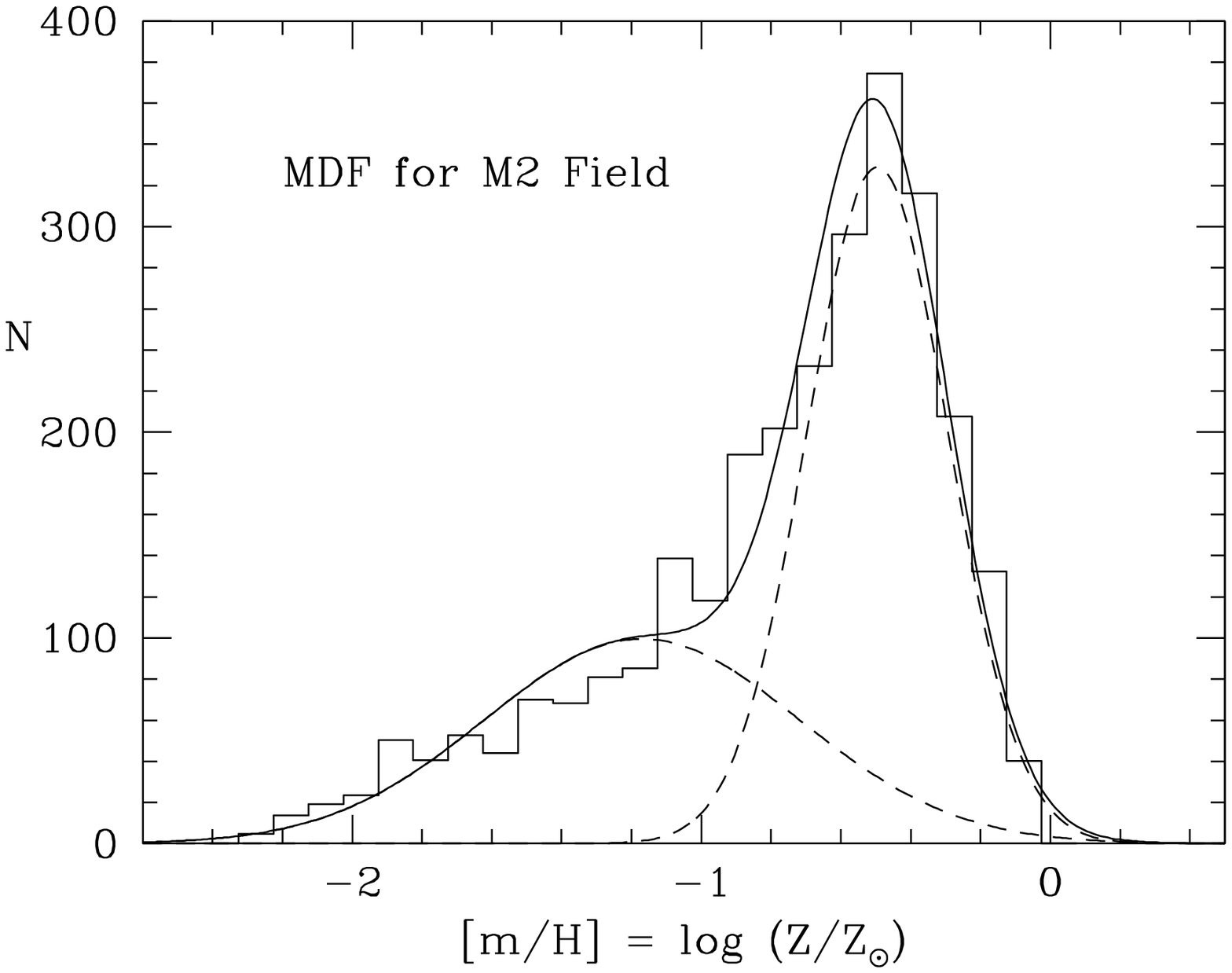}
\figcaption[fig11.eps]{M31 MDF (for $20.6 < I < 22.5$) with best-fitting Gaussians 
(dashed lines) and the combination of both (solid line).   \label{fig11}}

\includegraphics[width=6.5truein]{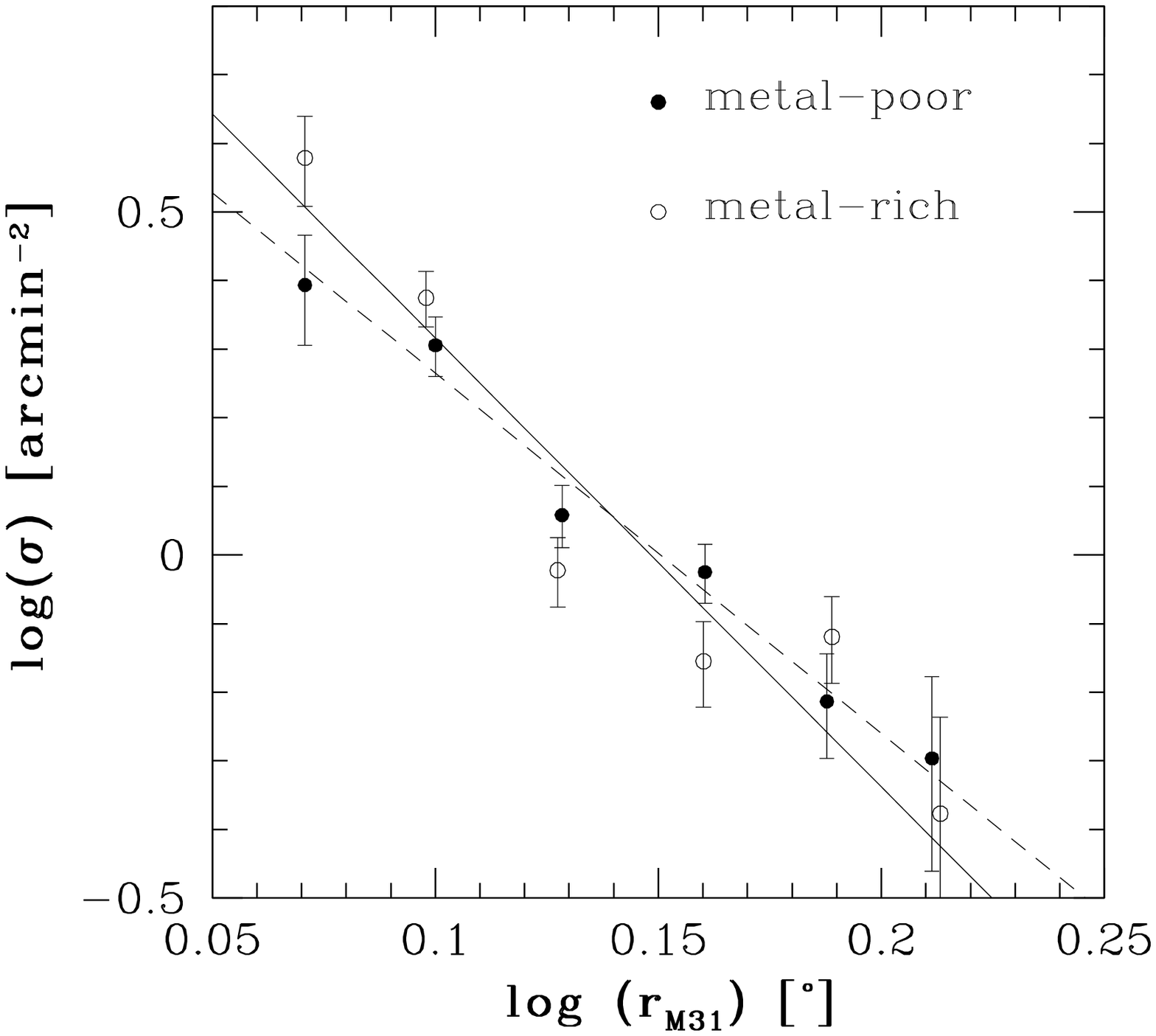}
\figcaption[fig12.eps]{Log-log plot of the number density of stars in the MDF with $20.6 < I < 22.5$, 
as a function of radius from M31.   Open circles denote those stars with $-2.5 <$[m/H]$<-1.0$ (metal-poor), 
the the filled circles are the metal-rich stars with $-0.6 <$ [m/H] $<-0.3$.    The metal-poor profile has been 
shifted upwards by a factor $\Delta$log($\sigma$)$=0.12$ to match the total number of stars in the metal-rich 
profile.  
The solid line is the best least-squares fit to the metal-rich profile ($\gamma = -6.54 \pm 0.59$), while the dashed 
line represents the best fit to the metal-poor profile ($\gamma = -5.25\pm 0.63$).  \label{fig12}}

\includegraphics[width=6.5truein]{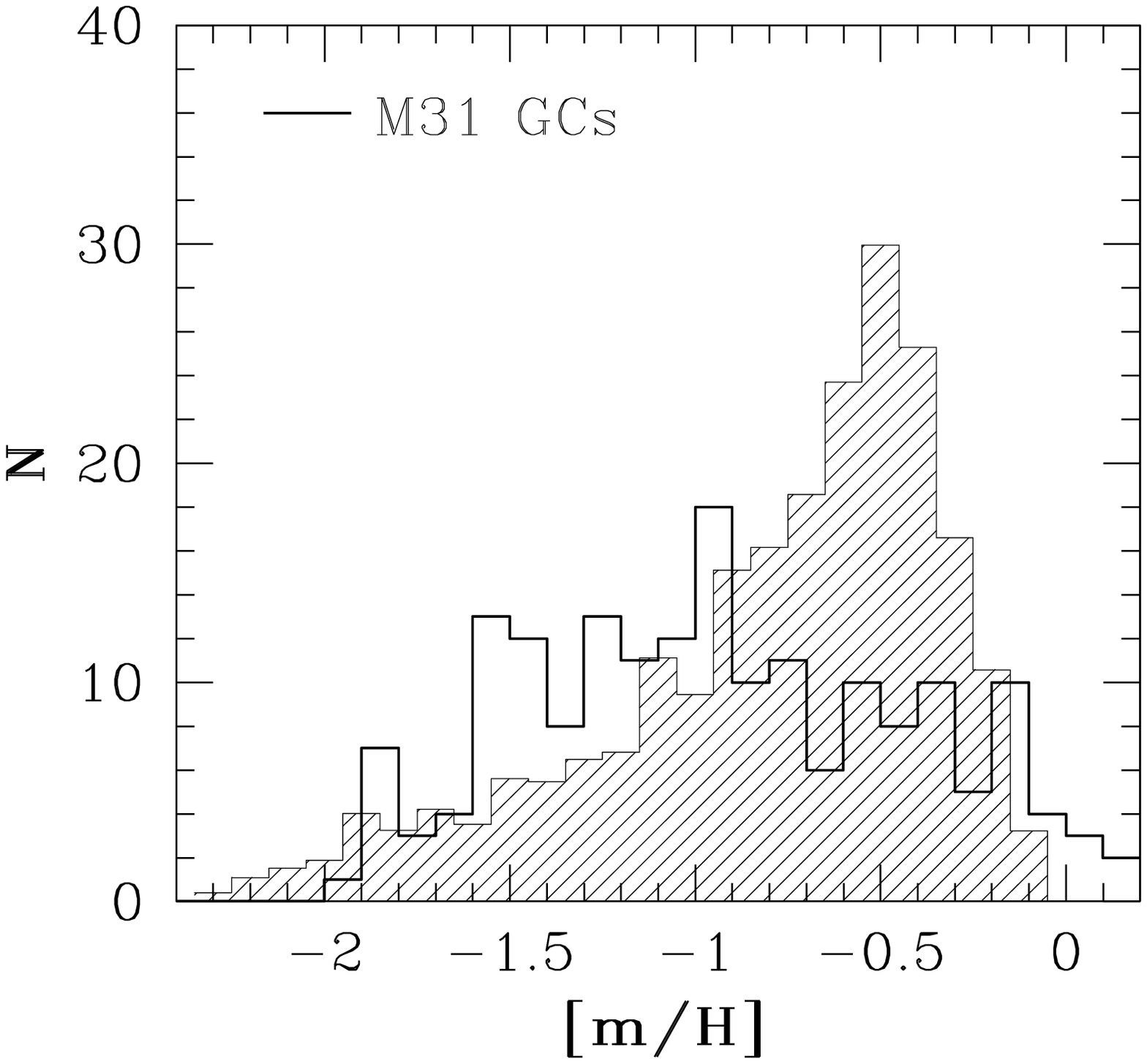}
\figcaption[fig13.eps]{Comparison between the MDFs for the M31 halo field (hatched region) 
and the M31 globular clusters with spectroscopically-derived metallicities from Barmby et al.(2000).  
[m/H] = [Fe/H] $+$ 0.3 has been applied to the GC metallicities. \label{fig13}}

\includegraphics[width=6.5truein]{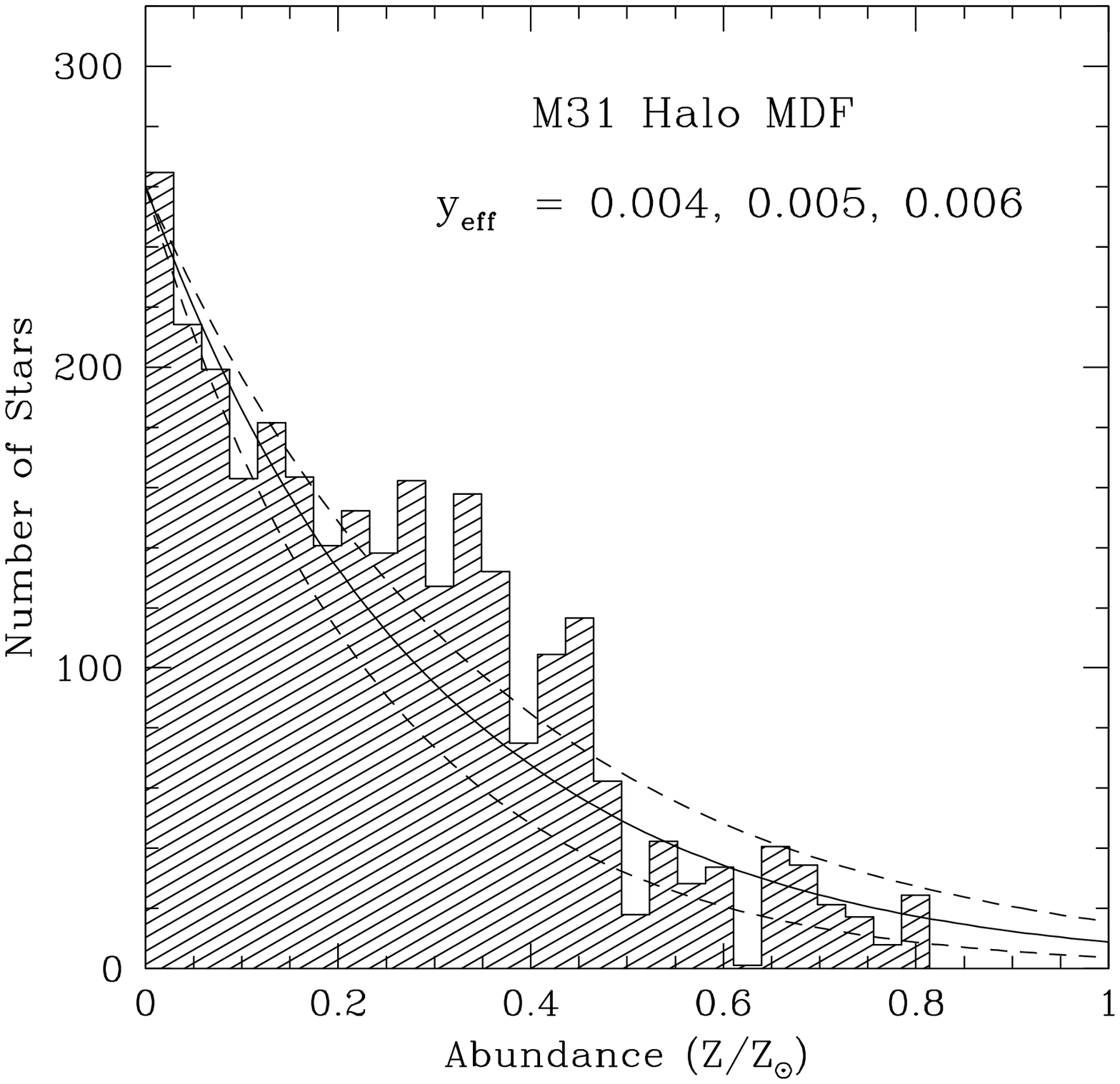}
\figcaption[fig14.eps]{Linear metallicity distribution function of the number of 
stars as a function of metal-abundance $Z$.  The three lines denote 
`leaky-box' chemical evolution models with 
$y_{eff} = 0.004, 0.005, 0.006$, with $Z_o = 0$ and $Z_{now}  = Z_{\odot}$.  
(The $y=0.005$ model is the solid line, with the other two as dashed
lines.) \label{fig14}}

\includegraphics[width=6.5truein]{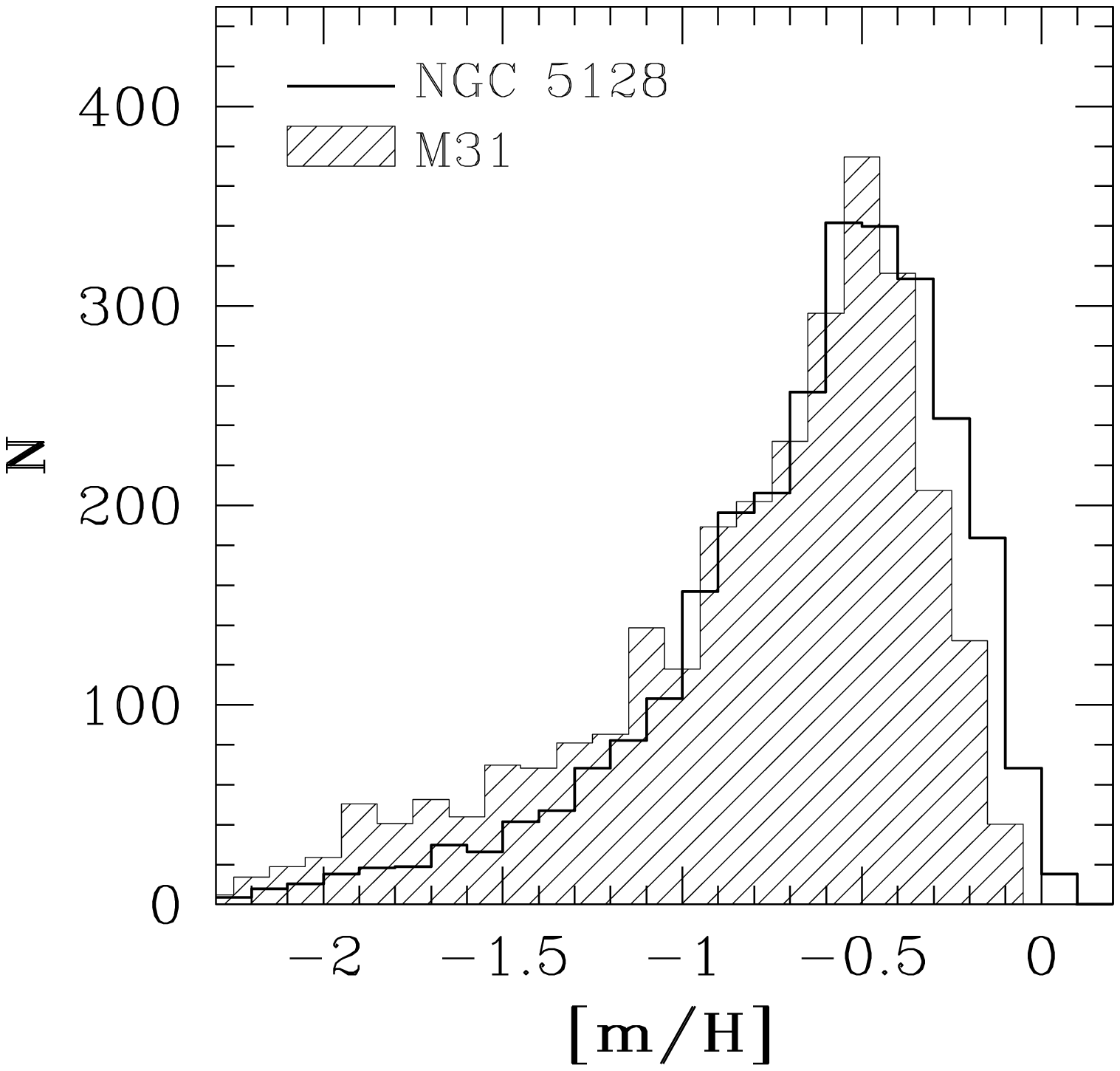}
\figcaption[fig15.eps]{Comparison of the M31 halo MDF with the outer halo MDF for NGC 5128 
(Harris et al. 1999; Harris \& Harris 2000).  The NGC 5128 data has been scaled by a factor 0.61 to normalize to the 
total number of total objects in both MDFs. \label{fig15}}

\end{document}